\documentclass{aa}
\usepackage{natbib,psfig,graphicx}
\usepackage{txfonts}
\usepackage{longtable,lscape}
\usepackage{rotating}
\usepackage{soul,color}
\def\mathnew{\mathsurround=0pt}
\def\simov#1#2{\lower .5pt\vbox{\baselineskip0pt \lineskip-.5pt
\ialign{$\mathnew#1\hfil##\hfil$\crcr#2\crcr\sim\crcr}}}

\def\MeV{Me\kern-0.11em V}
\def\keV{ke\kern-0.11em V}

\def\dd{deg$^{2}$}

\begin{document}

\title{The XMM-LSS survey: optical assessment and properties of different
X-ray selected cluster classes
\thanks{Based on observations obtained with MegaPrime/MegaCam, a joint
project of CFHT and CEA/DAPNIA, at the Canada-France-Hawaii Telescope
(CFHT) which is operated by the National Research Council (NRC) of
Canada, the Institut National des Science de l'Univers of the Centre
National de la Recherche Scientifique (CNRS) of France, and the
University of Hawaii. This work is based in part on data products
produced at TERAPIX and the Canadian Astronomy Data Centre as part of
the Canada-France-Hawaii Telescope Legacy Survey, a collaborative
project of NRC and CNRS. This work is also based on observations collected 
at TNG (La Palma, Spain), Magellan (Chile), and at ESO Telescopes at the La Silla 
and Paranal Observatories under programmes ID 072.A-0312, 074.A-0476, 076.A-0509,
070.A-0283, 072.A-0104, and 074.A-0360.}}

\offprints{C. Adami \email{christophe.adami@oamp.fr}}

\author{C. Adami\inst{1}
\and 
A. Mazure\inst{1} 
\and 
M. Pierre\inst{2} 
\and 
P.G. Sprimont\inst{3}
\and 
C. Libbrecht\inst{2} 
\and 
F. Pacaud\inst{2,8} 
\and 
N. Clerc\inst{2} 
\and 
T. Sadibekova\inst{3} 
\and 
J. Surdej\inst{3} \\
\and 
B. Altieri\inst{4}     
\and 
P.A. Duc\inst{2} 
\and 
G. Galaz\inst{7}  
\and 
A. Gueguen\inst{2}
\and
L.~Guennou\inst{1}
\and 
G. Hertling\inst{7} 
\and
O.~Ilbert\inst{1}
\and 
J.P.~LeF\`evre\inst{14}
\and
H. Quintana\inst{7} 
\and 
I. Valtchanov\inst{4}
\and 
J.P. Willis\inst{9} \\
\and 
M. Akiyama\inst{12}
\and
H. Aussel\inst{2}
\and
L. Chiappetti\inst{10} 
\and 
A. Detal\inst{3}          
\and
B.~Garilli\inst{10}
\and
V.~LeBrun\inst{1}
\and
O.~LeF\`evre\inst{1}
\and 
D. Maccagni\inst{10}
\and 
J.B. Melin\inst{13}
\and 
T.J. Ponman\inst{11}
\and
D. Ricci\inst{3}
\and
L.~Tresse\inst{1}\\
}

\institute{
LAM, OAMP, P\^ole de l'Etoile Site Ch\^ateau-Gombert 38, Rue Fr\'edr\'eric
Juliot-Curie,  13388 Marseille, Cedex 13, France
\and
Laboratoire AIM, CEA/DSM/IRFU/Sap, CEA-Saclay, F-91191 Gif-sur-Yvette Cedex, France
\and
Institut d'Astrophysique et de G\'eophysique, Universit\'e de Li\`ege,
All\'ee du 6 Ao\^ut, 17, B5C, 4000 Sart Tilman, Belgium
\and
ESA, Villafranca del Castillo, Spain
\and
UPMC Universit\'e Paris 06, UMR~7095, Institut d'Astrophysique de Paris,
F-75014, Paris, France
\and
CNRS, UMR~7095, Institut d'Astrophysique de Paris, F-75014, Paris, France
\and
Departamento de Astronom\'ia y Astrof\'isica, Pontificia Universidad
Cat\'olica de Chile, Casilla 306, Santiago 22, Chile
\and
Argelander-Institut f\"ur Astronomie, University of Bonn, Auf dem H\"ugel 71,
53121 Bonn, Germany
\and
Department of Physics and Astronomy, University of Victoria, Elliot Building,
3800 Finnerty Road, Victoria, V8V 1A1, BC, Canada
\and
INAF-IASF Milano, via Bassini 15, I-20133 Milano, Italy
\and
School of Physics and Astronomy, University of Birmingham, Edgbaston, Birmingham, B15 2TT, UK
\and
Astronomical Institute, Tohoku University
6-3 Aramaki, Aoba-ku, Sendai, 980-8578, Japan
\and
CEA/DSM/IRFU/SPP, CEA Saclay, F-91191 Gif-sur-Yvette, France.
\and
CEA/DSM/IRFU/SEDI, CEA Saclay, F-91191 Gif-sur-Yvette, France
}

\date{Accepted . Received ; Draft printed: \today}

\authorrunning{Adami et al.}

\titlerunning{Optical assessment and comparative study
of the C1, C2, and C3 cluster classes}

\abstract
{XMM and Chandra opened a new area for the study of clusters of galaxies. Not
only for cluster physics but also, for the detection of faint and distant
clusters that were inaccessible with previous missions.}
{This article
presents 66 spectroscopically confirmed clusters (0.05$\leq$z$\leq$1.5) within an
area of 6 deg2 enclosed in the XMM-LSS survey. Almost two thirds have been confirmed with
dedicated spectroscopy only and 10$\%$ have been confirmed with
dedicated spectroscopy supplemented by literature redshifts.}
{Sub-samples, or classes,  of
extended-sources are defined in a two-dimensional X-ray parameter space
allowing for various degrees of completeness and contamination. We describe
the procedure developed to assess the reality of these cluster candidates
using the CFHTLS photometric data and spectroscopic information from our own
follow-up campaigns.}
{Most of these objects are low mass clusters, hence
constituting a still poorly studied population. In a second step, we
quantify correlations between the optical properties such as richness or
velocity dispersion and the cluster X-ray luminosities. We examine the
relation of the clusters to the cosmic web. Finally, we review peculiar
structures in the surveyed area like very distant clusters and fossil
groups.
}
{}  


\maketitle

\section{Introduction}

With the quest for the characterisation of the Dark Energy properties and the 
upcoming increasingly large instruments (JWST, ALMA, LSST, EUCLID, etc. ...) 
the beginning of the 21st century is to be an exciting time  for cosmology. In 
this respect, a new era was already open for X-ray astronomy by the XMM-Newton 
and Chandra observatories in 1999. The increasing amount of high quality 
multi-wavelength observations along with the concept of  ``multi-probe'' approach 
is expected to provide strong constraints on the cosmological models. In this 
context,  X-ray surveys have an important role to play, as it was already the 
case in the 80s and 90s (e.g. Romer et al. 1994, Castander et al. 1995, Collins 
et al. 1997, Henry et al. 1997, Bohringer et al. 1998, Ebeling et al. 1998, Jones 
et al. 1998, Rosati et al. 1998, Vikhlinin et al. 1998, De Grandi et al. 1999, 
Romer et al. 2000, and ref. therein). New cluster surveys are presently coming to 
birth (e.g. Romer et al. 2001,  Pierre et al. 2004, Finoguenov et al 2007).  \\
 
One of them, the XMM-LSS survey, covers 11 \dd\ at a sensitivity of $\sim$
$10^{-14}$ erg/s/cm$^{2}$ at 0.5-2keV for spatially-extended X-ray
sources and is currently the largest contiguous deep XMM cluster survey. 
This sky region is covered by parallel surveys in multiple complementary wavebands
ranging from radio to the $\gamma$-ray wavelengths (Pierre et al., 2004)
and therefore constitutes a unique area for pioneering studies. It can
for instance detect a Coma-like cluster at z$\sim$2.
A number of articles describing the properties of the XMM-LSS source population 
have been published by e.g. Pierre et al (2006) and Pacaud et al (2007) for clusters 
of galaxies and Gandhi et al (2006) for AGNs; the complete X-ray source catalog along 
with optical identifications for the first 5 \dd\ of the survey was published by Pierre et al (2007). 

One of the major goals of the XMM-LSS survey is to provide samples of galaxy clusters with well 
defined selection criteria, in order to enable cosmological studies out to redshift $z \sim 1.5$. 
Indeed, monitoring selection effects is mandatory not only to study  the evolution of the cluster 
X-ray luminosity (i.e. mass) function or of the 3-D cluster distribution but also, as shown 
by Pacaud et al. (2007), to characterise the evolution of the cluster scaling laws such as the 
luminosity-temperature relation. We have put special emphasis on the X-ray selection 
criteria in the XMM-LSS survey. The procedure enables the construction of samples having various 
degrees of completeness and allows for given rates of contamination by non cluster sources.  
The subsequent optical spectroscopic observations constitute the ultimate assessment of the 
clusters, thus operates the purification of the samples. 

In a first paper, Pacaud et al (2007) presented the Class One (C1) clusters pertaining to 
the first 5  \dd\ of the survey (the ones with the highest apriori probability to be real 
clusters). The C1 selection yields a purely X-ray selected cluster 
sample with an extremely low contamination level and corresponds to rather high surface 
brightness objects. The present article summarises these former findings including now  
the clusters selected from less stringent X-ray criteria (C2 and C3) and including the 
contiguous Subaru Deep Survey (SXDS, e.g. Ueda et al. 2008).  
The C2 and C3 objects  presented here come from an initial sample with a higher 
degree of contamination, but  have all passed the final spectroscopic tests.  Compared  to the C1 
clusters, they are fainter and correspond a-priori to less massive clusters or to groups at a redshift 
of $\sim$0.5: this is a population that is for the first time systematically unveiled by the 
XMM-LSS survey. A few massive very distant clusters are falling into this category too.  

The present study is the first attempt to give a comprehensive census (X-ray and optical properties) 
of the low-mass cluster population within the $0<z<1$ range. Search for correlations between 
optical and X-ray properties has already been a long story from, e.g, Smith et al. (1979) or 
Quintana $\&$ Melnick (1982). However, with more than 60 
spectroscopically confirmed clusters, the current sample constitutes, by far, the spectroscopically 
confirmed cluster sample with the highest surface density ever published.
The article is organised as follows. Next section describes 
the X-ray cluster selection. Section 3 presents the available optical photometric and 
spectroscopic data. Section 4 explains the adopted cluster validation procedure, the new 
X-ray luminosity computations, and presents the resulting catalog.  Then, the global 
properties of each cluster class and category are examined in Section 5 and, 
subsequently,  the properties of the cluster galaxy population in Section 6. Section 7 
details the z=1.53 cluster and investigates possible peculiar structures in the survey. 
Finally Section 8 gathers the conclusions. The 
two appendixes discuss the accuracy of photometric redshifts in the context of dense environments
and lists additional redshift structures found in the course of the study. 

Throughout the paper we assume H$_0$ = 71 km s$^{-1}$ Mpc$^{-1}$, $\Omega _m$=0.27, and
$\Omega _{\Lambda}$=0.73 (Dunkley et al 2009). All magnitudes are in the $AB$ system.

\section{The initial cluster candidate selection}

The clusters presented in this paper are for the great majority X-ray selected.
The XMM-LSS pipeline (Pacaud et al 2006) provides for each detected source some 20 parameters 
(co-ordinates, count rate, etc..). Out of these, two are especially relevant for the characterisation 
of extended sources: the extent measurement ({\tt EXT}) and the likelihood of extent ({\tt EXT\_LH}). 
We recall (as defined by Pacaud et al 2006) that the "extent" parameter
is the core radius of the beta-profile fit by the survey pipeline to each
source, assuming a fixed beta of 2/3. The 
cluster selection basically operates in this two-dimensional space and has been extensively 
adjusted and tested using simulations of hundreds of XMM images. This allows the definition of 
three cluster samples.
\begin{itemize}
\item The C1 class is defined such that $\sim$ no point sources  are misclassified as extended 
(i.e. less than 1\% of the cluster candidates are point sources) and is described by {\tt EXT} 
$> 5'',$ {\tt EXT\_LH} $> 33$ plus an additional boundary on the detection likelihood, set 
to be greater than 32.
\item The C2 class is limited by  {\tt EXT} $> 5'',$ {\tt EXT\_LH} $> 15$  and  displays an
a-priori contamination rate of about 50\%.
\item The C3 clusters are faint objects and thus, have less-well characterized X-ray properties. They may
be located at the very edge of the XMM field of view or suffer contamination by point sources. They 
therefore result from a subjective selection mostly based on a visual 
inspection of the X-ray and optical data; their selection function is up to now undefined.
\end{itemize}
More details about the classification can be found in Pacaud et al (2006) and Pierre et al (2006).\\

In this paper, we have presented a large sample of X-ray clusters, including the 29 C1 confirmed 
clusters published by Pacaud et al. (2007). These C1 
clusters were already unambiguously confirmed, but we take the occasion of this publication to reprocess the 
associated optical spectroscopic data following the standard method developed in the present paper. 
This will provide a unique homogeneous cluster sample. The clusters pertaining to this paper are, 
for most of them, located in the first 5 \dd\ of the XMM-LSS region, supplemented by the Subaru  Deep Survey. 
The validated C1, C2, and C3 samples are presented 
in Tables~\ref{tab:cand1C1}, ~\ref{tab:cand1C2}, and ~\ref{tab:cand1C3}. In these Tables, \textbf{XLSS} 
catalog names refer to sources published in Pierre et al. (2007). \textbf{XLSSU} catalog names refer 
to sources whose fields were not yet considered in XLSS (for example flagged bad or in SXDS fields) and 
reobserved (or reprocessed later), or which were below the detection likelihood threshold in the input
data set used as source for XLSS.

In the course of the data inspection, we have also identified a few clusters using optically based 
criteria such as the red sequence or the gapper method. Our spectroscopic data set allowed us to confirm 
them as bona fide clusters, although these objects are not detected in the X-rays by the current 
version of the XMM-LSS pipeline or the association between X-ray detected sources and optical clusters
is not straightforward. We thought of interest to publish these objects and they are listed in Tables
~\ref{tab:cand1C0} and ~\ref{tab:cand2}.

We now describe the involved optical data and the general identification processes.

\section{The optical data}

\subsection{The optical spectroscopic data}

\begin{figure}
\centering
\mbox{\psfig{figure=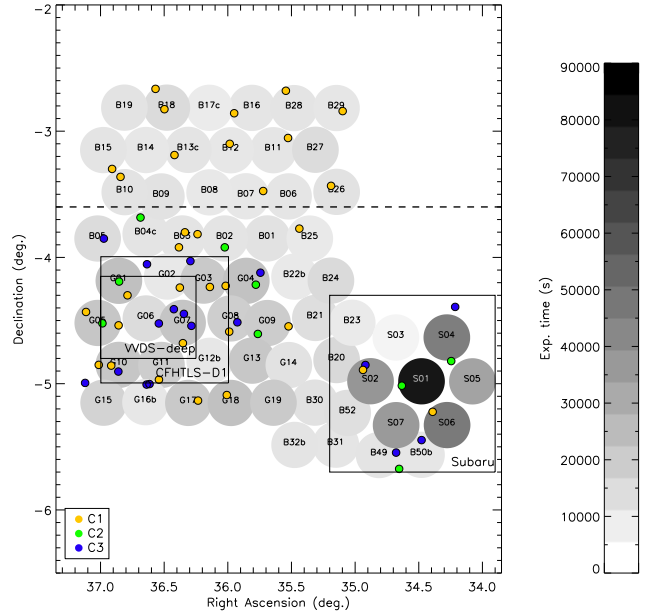,width=9cm,angle=0}}
\caption[]{Map showing the different involved surveys. The grey level disks are the 11' central areas of the XMM pointings 
(exposure time depends on the greyness). Large squares show the spectroscopic VVDS-deep and Subaru Deep surveys, and
the CFHTLS D1 field. C1, C2, and C3 clusters are also shown. Above a 
declination of -3.6deg, only g',r',z' coverage is available, hence no
photometric redshifts are derived for this zone.}
\label{fig:figpres}
\end{figure}

We have been performing a dedicated spectroscopic follow-up of all C1 clusters and of a number of 
C2 and C3 clusters. These PI observations are listed in Table~\ref{tab:spectro} and provide about 
2000 redshifts to date. We supplemented this data set with the VVDS deep (e.g. LeF\`evre et al. 2005: 
$\sim$11000 redshifts in 0.49 deg$^2$) and ultradeep (LeF\`evre et al. in preparation) data, and with 
a redshift compilation pertaining to the Subaru Deep Survey (Ueda et al 2008) included in the 
XMM-LSS area. Some 200 other redshifts were also available from  NED for part of the area.  
We show in Fig.~\ref{fig:figpres} the location of these different surveys, as well as the 
exposure time of the different XMM fields.

\begin{table}
\caption{PI spectroscopic runs involved in the present paper.}
\begin{center}
\begin{tabular}{ccccl}
\hline
Telescope & Instrument & Year & Nights & Run ID  \\
\hline
Magellan & LDSS2  & 2002 & 2 & - \\
Magellan & LDSS2  & 2003 & 4 & - \\
NTT & EMMI  & 2003 & 3 & 72.A-0312\\
NTT & EMMI  & 2004 & 4& 74.A-0476\\
NTT & EMMI  & 2005 & 3 & 76.A-0509\\
TNG & DOLORES  & 2007 & 4& AOT16/CAT\_75 \\
VLT & FORS2  & 2002 & 3 &  70.A-0283\\
VLT & FORS2  & 2003 & 4 & 72.A-0104\\
VLT & FORS2  & 2004 & 4.5 & 74.A-0360 \\
\hline
\end{tabular}
\end{center}
\label{tab:spectro}
\end{table}

Individual redshift measurements of spectra resulting from the PI data were made 
following a procedure  similar to that adopted by the VVDS survey. Each spectrum was 
independently measured by several people and the redshift subsequently validated by 
a moderator. Quality flags were assigned to each measurement following the VVDS rules: 
flag 0 indicates an inconclusive result, flag 1 means a probability of 50$\%$ that 
the assigned redshift is wrong, flag 2 means a probability of 25$\%$,  flag 3 
means a probability of 5$\%$,  flag 4 means a probability of 1$\%$, and  flag 9 
means we have assigned a redshift with a single line using absent lines in
order to limit the possibilities. These percentage levels proved to be
reliable in the VVDS survey (LeF\`evre et al. 2005).

Our spectroscopic redshifts having quite heterogeneous origins (different
telescopes, instrumentations, and resolutions), it is therefore useful to
compute the ability to measure a redshift and the achieved velocity resolution. In order to achieve 
such a goal, we chose to compare the PI data to the VVDS survey, which provides a 
well qualified set of data. Only 26 galaxies both measured by the VVDS and our 
dedicated follow-up have a quality flag greater or equal than 2. For these objects, given 
the VVDS quality flags (6 flags 2, 6 flags 3, and 14 flags 4), we expect to have 3.2 wrong
redshifts. We indeed find 3 redshifts differing by more than 0.05 between the
PI and VVDS data. VVDS spectroscopic redshifts are expected to have a typical uncertainty of 280
km/s (from repeated VVDS redshift measurements, Le F\`evre et
al. 2005). Excluding all redshifts with differences greater than 0.02, we find
a typical uncertainty between PI and VVDS redshifts of  340 km/s.
Even with a comparison done on a rather limited
size sample,  the PI redshifts appear thus reliable in the [0.,1.]
redshift range and in the [18,23] I VVDS magnitude range.

Finally, it has to be mentioned that, for the spectroscopic sample, no completeness, 
neither spatial nor in luminosity, can be globally defined because of the various data
origins.

\subsection{The optical photometric data}

Most of the XMM-LSS area is covered by the Canada-France-Hawaii Telescope Legacy Wide Survey
(CFHTLS-Wide\footnote{http://terapix.iap.fr/cplt/T0006/T0006-doc.pdf}). This survey, performed by 
means of the MegaCam camera,  covers some 171 deg$^2$  in 4 independent patches with five filters 
($u*,g',r',i'$ and $z'$). Resulting catalogs are 80\% complete down to $i'_{AB}$=24. The Wide 
survey encloses a sample of about 20$\times$10$^6$ galaxies inside a volume size of $\sim 1$ Gpc$^3$, 
with a median redshift of z$ \sim 0.92$ (Coupon et al 2009). Northern of Dec= -3.6 deg, the 
CFHTLS data were complemented by PI MegaCam observations (3 \dd) performed in {\em g', r', z'} 
at the same depth as the CFHTLS; they were reduced following the same procedure. 

The optical images and catalogs were primarily used to check for the presence of galaxy concentrations 
coinciding with the extended X-ray emission. The CFHTLS data (only the T0004 release was 
available at the beginning of the present study) enabled the determination of photometric 
redshifts in the best fit template (Coupon et al. 2009). These 
photometric redshifts cover 35 deg$^2$ in the T0004 partially 
overlapping with the XMM-LSS area. They were computed using a template-fitting method, calibrated 
with public spectroscopic catalogs. The method includes correction of magnitude systematic offsets. 
The achieved photometric redshift precision $\sigma _z$/(1+z) is of the order of 
0.04 with a catastrophic error percentage of less than 5$\%$ at i'$\leq$23 (the magnitude limit we 
adopted for the photometric redshifts).

\section{"Cluster candidate" validation process}
 
\subsection{General method}

 Extragalactic extended X-ray emission is the signature of a deep
 gravitational potential well. Apart from the hypothetical  "dark clusters", 
this potential well coincides with a galaxy over-density. The system (cluster or group)  
is therefore  detectable using optical information only.  In this article, we aim at
 assessing the presence of optical structures corresponding to the X-ray cluster candidates. 
Such systems are expected to  manifest themselves as compact structures in redshift space 
(both spectroscopic and photometric ones) and as localized excess in projected galaxy density
 maps.  
 
To perform such an analysis, we make use of the  two optical data sets mentioned above.
The investigated lines of sight (centered on the X-ray emissions)
were initially selected from the condition that at least two spectroscopic redshift measurements
(whatever their values) are available within the X-ray isophotes. The subsequent conditions
were more stringent depending on the cluster nature (see below).

The CFHTLS Wide survey and subsequent analyses (e.g. Coupon et al. 2009) provide us with 
galaxy positions as well as their  
apparent and absolute magnitudes, photometric redshifts and the corresponding  "galaxy
types"  $T$ (from the spectral fitting performed during the photometric
redshift computation). With the exception of the usual "masking problems" due to bright
stars or CCD defaults, photometric data are homogeneous and allow us to define
complete sub-samples in terms of spatial extension or in magnitudes. 
Limitations to these data are the redshift range within which photometric
redshifts are reliable and the adopted magnitude limit. Here we restrict
ourselves to $0.2 <  z < 1.2$ and $i' = 23$ (see Coupon et al. 2009). This
limiting magnitude  will affect partly the use of photometric data  to detect
structures. Indeed, the characteristic magnitude m* of the Schechter
luminosity function is about $i' = 20$ at $z = 0.5$ and $i' = 22.5$ at $z = 1$
leading to sampled luminosity function ranges of about m*+ 3 to m* + 0.5 at 
these respective redshifts.
One drawback is  therefore that  for $z > 0.5$ the number of galaxies actually
belonging to a structure will be rapidly overcome by the background
contamination (see e.g Table 1 of Adami  et al 2005). One way to fight this
contamination will be to use redshift slices defined on a photometric redshift
basis (see Mazure et al. 2007) but the
range covered in magnitude by structure members will remain limited.

In order not to bias  the optical characterisation of
the X-ray sources, the information concerning the C1, C2,
C3 classification was used only at the very final stage.

\subsection{Different analysis steps}

The  first step concerns the expected compactness in spectroscopic redshift space. To 
reveal such compact associations, we used the already well tested and used "gap
method" (e.g. Biviano et al 1997, Rizzo et al 2004). It looks for significant gaps
between successive galaxy velocities within the ordered redshift distribution
obtained along a given line of sight. As in Adami et al (2005), we use a gap defined by
$g$ = 600 (1+z) km/s which was optimum for the considered redshift
range. When the velocity difference between 2 successive galaxies is smaller
than $g$, they are assigned to belong to a common structure, otherwise they
are put in different groups. 

Since the lines of sight most of the time sample redshifts up to at least z = 1, this first
step of the analysis ends in general with several groups. Thus, with the mean
redshift of every group, a  cosmological distance was assigned, a physical
region of 500 kpc (radius) defined and the galaxies within  this radius are selected
as potential real cluster members. We choose this size as being representative of clusters
in terms of membership of galaxies w.r.t the field. Taking larger regions
would decrease any real contrast, while taking smaller regions would decrease the
number of true members.  As a second step we then apply the usual ROSTAT tools
(Beers et al. 1990) on individual redshift groups to test for final membership 
and definition of the group properties (robust redshift locations and scales 
with their corresponding bootstrap errors).

As already mentioned, several groups are in general identified along the
lines of sight. Before comparing the galaxy distribution and the X-ray isophotes, we
used then, when available, the CFHTLS photometric redshift information. As a third
step, we selected galaxies in photometric redshift slices (of width: $\pm$
0.04 (1+z), see Coupon et al. 2009) around the mean redshift of the considered  
group and produced iso-contours of numerical galaxy density  (see  Mazure et al  
2007 for details and previous application). It is expected that the optical 
group physically associated with the X-ray emission will show up with a clear
density contrast located next to the position of the X-ray center. This is
because the use of photometric redshift slices removes a large part of the fore and
background contaminations. We also look, as another check,  at the
photometric redshift distribution  within various central regions compared to
the one in the largest available region, conveniently renormalized and defined
as the  "field". Again, one expects a clear contrast at the redshift values
given by the spectroscopy.

\begin{figure}
\centering
\mbox{\psfig{figure=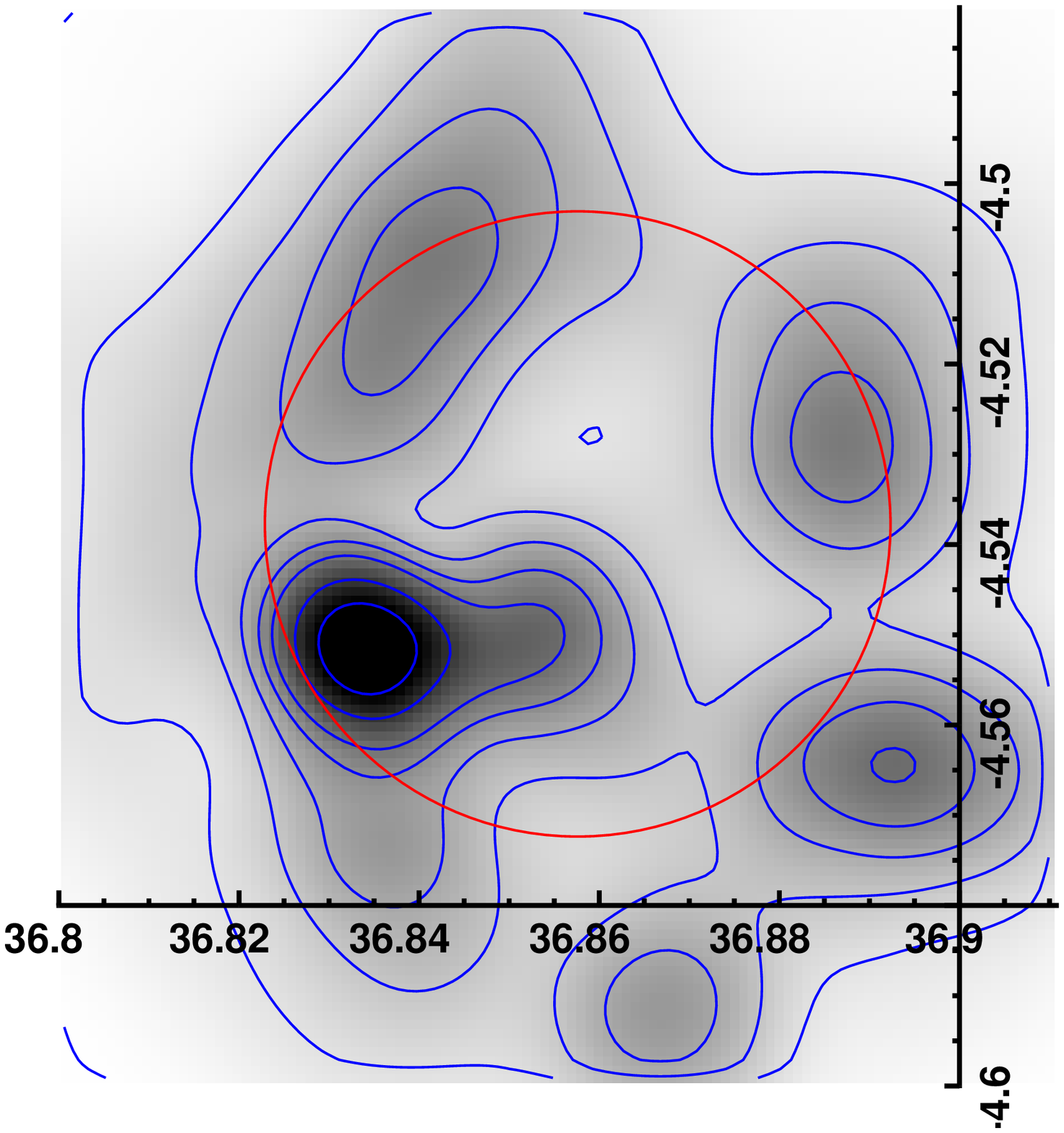,width=7.5cm,angle=0}}
\mbox{\psfig{figure=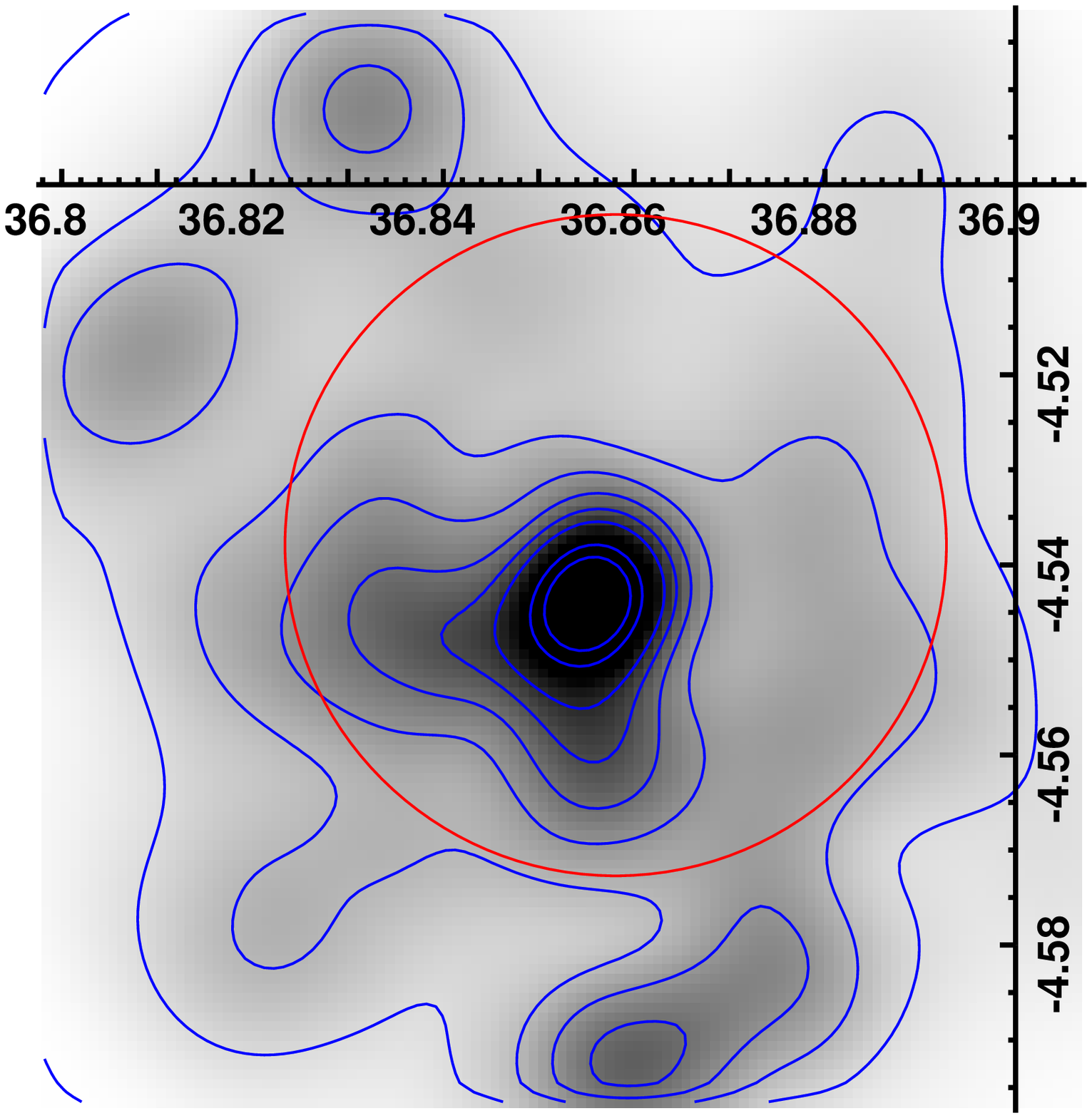,width=7.5cm,angle=0}}
\mbox{\psfig{figure=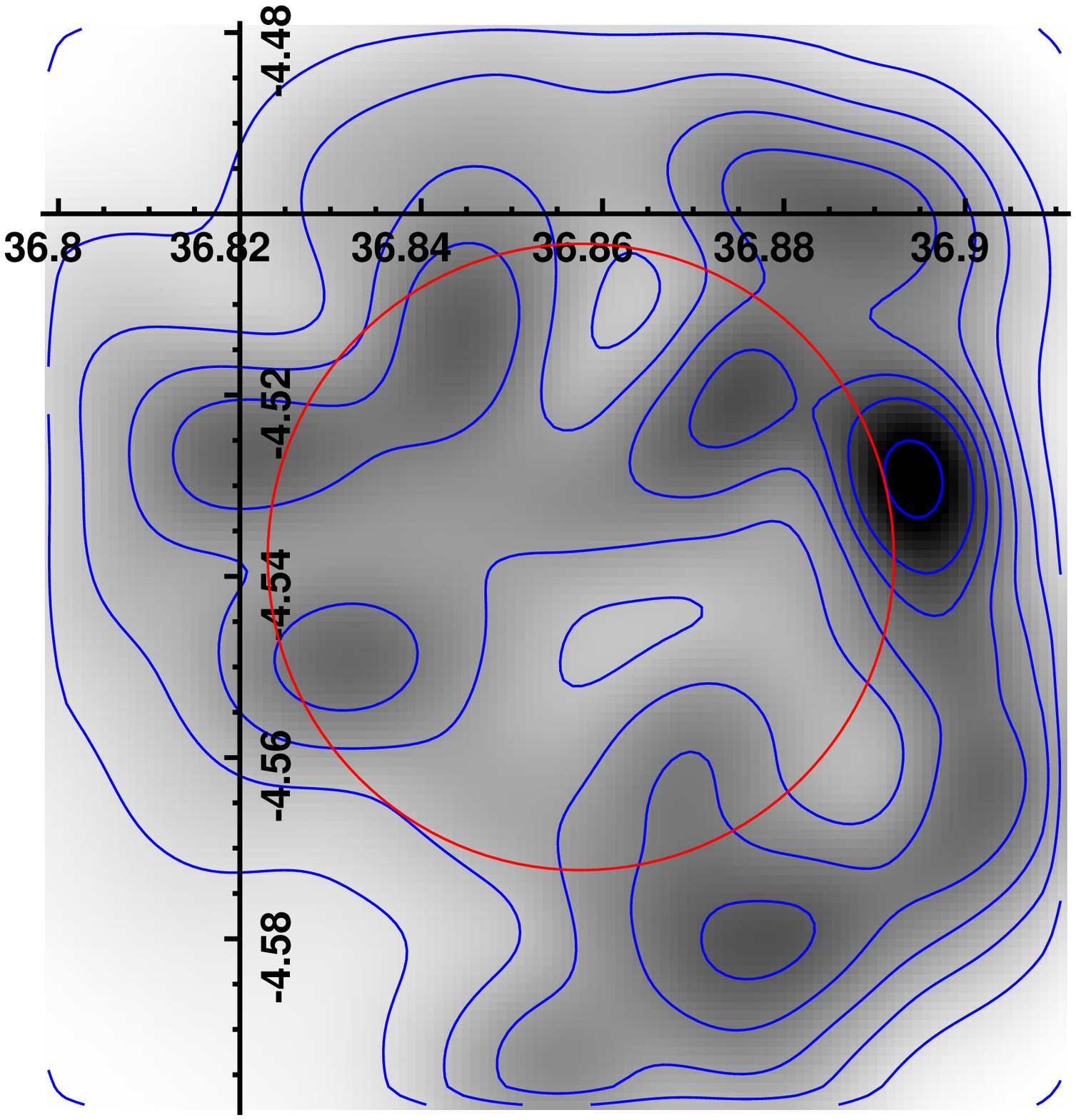,width=7.5cm,angle=0}}
\caption[]{Isodensity maps of the numerical density of galaxies within photometric
  slices of width $\pm$ 0.04 (1+z) around the group redshifts. From top to bottom: 
z $\sim$ 0.2, z $\sim$ 0.3, z $\sim$ 0.6. The best
  agreement with the X-ray emission of XLSSC 013 is obtained at  z = 0.3. Large red circles 
are the same as in Fig.~\ref{fig:fig3}.}
\label{fig:fig2}
\end{figure}

An illustration is given with the source XLSSC 013 in the XMM-LSS
database. Three main groups were identified along the line of sight (z $\sim 0.2$ with
9 redshifts, z$\sim0.3$ with 26 redshifts, z $\sim 0.6$ with 5
redshifts). Consecutive examination of both the photometric redshift
distribution and the numerical density histograms gives strong evidence for
the z$\sim$0.3 group to be chosen (see Figs.~\ref{fig:fig2} and ~\ref{fig:fig3}).

However, as mentioned above, photometric redshift data were not always available and
spectroscopic data could be very sparse (our velocity dispersion measurements are then subject
to very complex selection functions in the target selection when measuring and collecting 
galaxy redshifts). The final selection is then done by a
visual inspection of X-ray and optical maps taking into account all the
informations available. Fig.~\ref{fig:fig3} shows the group at z = 0.3 chosen for
XLSSC 013. As an extreme contrary case, we show in Fig.~\ref{fig:fig4} XLSSC 035 
for which only a few redshifts were available.  The fact that  a giant galaxy at z = 0.069 
lies at the center finally pleads in favor of that redshift (Fig.~\ref{fig:fig4}) in the present
paper. We note however that a z$\sim$0.17 galaxy layer is also detected along this line of sight
and we could deal with a superposition effect.

\begin{figure}
\centering
\caption[]{XMM-LSS X-ray contours for system XLSSC 013 with cluster member galaxies with a measured
  redshift (between z=0.3049 and 0.3112) superimposed. The red circle corresponds to a radius of 500 kpc
at z = 0.3.}
\label{fig:fig3}
\end{figure}

\begin{figure}
\centering
\caption[]{XMM-LSS X-ray contours for system XLSSC 035 with galaxies with measured
  redshifts superimposed. The red circle corresponds to a radius of 500 kpc at z = 0.069.}
\label{fig:fig4}
\end{figure}

\subsection{Results}

We examined 34 C1 candidate X-ray sources. Identification fails for only two
lines of sight mainly because very few redshifts were available in the X-ray
region and/or no photometric redshifts. All identified sources were classiﬁed
as galaxy clusters; this means that at least 95\% of the C1 objects are real
clusters (when obvious nearby galaxies - which show also a diffuse X-ray
emission - are excluded). Among the C2 and C3 candidates, only those having
2 redshifts within the X-ray isophotes were selected for the present
analysis. As our current spectroscopic data set is heterogeneous and does
not provide a systematic targeting of all C2 and C3 cluster candidates, it
is not possible to draw firm conclusions about the effective contamination
rate (in terms of non-cluster sources) for these populations.
We may only state that for all C2 (resp. C3) sources having yet at least 2
spectroscopic redshifts within the X-ray isophotes, more than 80$\%$ (resp. 50$\%$) of
the examined sources turned out to be real clusters. 

An additional potential X-ray source was also 
discovered (C555 in Table~\ref{tab:cand1C3}). Not listed in
Pierre et al. (2007), this source is merged with XLSSU J022533.8-042540. 
We detected a very clear associated galaxy structure in optical. A manual extraction
of the X-ray source gives a count rate of 0.003$\pm$0.001 counts per second 
([0.5-2keV]). 

For seven of the analysed lines of sight, the association between X-ray source 
and optical galaxy concentration was not obvious or the X-ray source was not significantly
different from the background. 
However, these clusters are identified on the basis of the color magnitude relation 
(for 2 of them) or are detected as significant galaxy overdensities in Adami et al. (2010)
using photometric redshifts during the analysis. All these objects have been classified as C0 clusters. 

C1, C2, C3, and C0 clusters are presented in Tables~\ref{tab:cand1C1}, ~\ref{tab:cand1C2},
~\ref{tab:cand1C3}, and ~\ref{tab:cand1C0}. Almost two thirds have been confirmed with
dedicated spectroscopy only and 10$\%$ have been confirmed with
dedicated spectroscopy supplemented by literature redshifts.

We compared the cluster redshifts listed in the present paper (see also
next section) with the estimates already published within the XMM-LSS framework (from 
Pacaud et al. 2007 and 
Bremer et al. 2006: 29 C1 clusters and 1 C2 cluster), and we found the expected good agreement.
This is not surprising as Pacaud et al. (2007) and Bremer et al. (2006)
are included in the presently used spectroscopic redshift sample. However, redshift
measurements have been redone on a more homogeneous basis and sometimes with new data.
The difference is only 0.00075$\pm$0.00329 when excluding XLSSC 035. For this cluster, we detected
a possible error in the individual redshifts measurement process. The
central galaxy seems to be at z=0.069 and not 0.17 as stated in Pacaud et al. (2007: cluster
redshift changed to z=0.069). We are in the process of acquiring more data
in order to definitively solve this case. We also note that 
the central galaxies of XLSSC 028 are also at z$\sim$0.3 and not at z$\sim$0.08 as
stated in Pacaud et al. (2007: cluster redshift unchanged at z$\sim$0.3).

\begin{figure}
\centering
\mbox{\psfig{figure=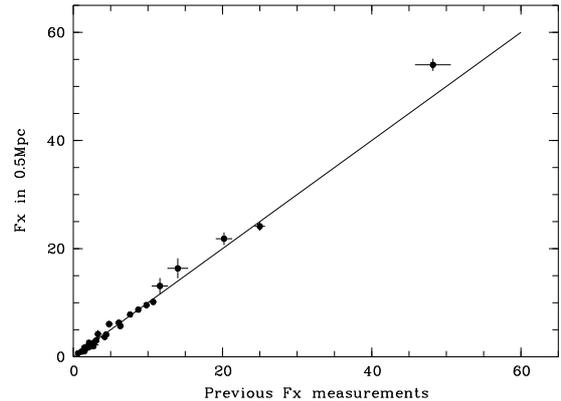,width=8cm,angle=270}}
\caption[]{Previous and present cluster flux (in 
a 0.5Mpc radius) comparisons.}
\label{fig:previous}
\end{figure}

The agreement is very good for the [0.5-2 keV] fluxes measured in a 500kpc radius
(Fig.~\ref{fig:previous}). 

For the Subaru Deep Survey region, we compared our detections with the extended X-ray source 
catalog of Ueda et al. (2008) and with the
structure catalog of Finoguenov et al. (2010). Nine of our X-ray clusters are inside the area covered
by these catalogs and six are also detected by these authors. Redshifts are always in very good agreement. 
Finoguenov et al. (2010) list in their paper 57 structures inside this area. However, their selection 
function (completeness/contamination) for the X-ray extended sources as well as the characteristics of 
these sources (extent-measurement along with error or likelihood) are not published, thus
preventing any meaningful comparison between the two samples. Moreover, as
shown by Pacaud et al (2006) a flux limit, unless it is set very high,
cannot define a complete uncontaminated sample of extended sources.

We finally performed a comparison with independently optically detected clusters in the 
literature. Limiting ourselves to studies giving a galaxy velocity dispersion 
estimate, we have five detections in common with Hamana et al. (2009: see 
Table~\ref{tab:cand1C1}). All these
clusters are C1 structures. Redshifts are always in good agreement. Galaxy velocity dispersions
are also consistent within error bars with an exception for XLSSC 050 where we find
408$\pm$96 km/s and where Hamana et al. (2009) find 739$_{-86}^{+150}$km/s. This structure
being very complex, the galaxy velocity dispersion is however very dependent
on the selected galaxies and on the exact center choice. 

 \subsection{Updated X-ray luminosities}

We apply the principle of "aperture photometry" to the flux measurement of the 
X-ray clusters, which avoids any other assumption than spherical symmetry as to 
the cluster shape. We note that Pacaud et al. (2007) used a beta-model fitting, which 
is not possible for the larger sample presented here, that comprises faint 
objects. For these C2 and C3 objects, having sometimes at most some hundred counts, it is not
possible to perform  a semi-interactive spatial fit as in Pacaud et al (2007),
i.e. letting the core radius and the beta value as free parameters. The resulting 
uncertainty would be very large.

We integrate the count rate in concentric annuli and derive the 
uncertainties by using Poisson statistic. Then, considering the count rate 
in each annulus, we stop the integration at the radius of the annulus 
for which the corresponding count-rate increase is comparable to the background 
1-sigma fluctuation. This program operates in semi-interactive mode leaving the 
possibility to optimize the determination of the X-ray centroid and of the background 
level. The measurement yields for each cluster the total MOS1 + MOS2 + PN count-rate 
within a radius 500 kpc. The fluxes have been obtained 
assuming a fixed conversion factor into the [0.5-2] keV band using a constant 
conversion factor of 9$\times$10$^{-13}$ [(ergs/cm2/s)/(cnts/s)]. This value was calculated 
using Xspec from an APEC emission model with the following parameters: z=0.5, 
T=2 keV, Nh=2.6$\times$10$^{20}$ cm$^{-2}$, Ab=0.3. Bolometric luminosities (also within a 
500kpc radius) listed in the tables 
were also calculated with Xspec from the measured fluxes using the Pacaud et al. 
(2007) and the Bremer et al. (2006) temperatures when available. We used redshifts described
in the next section. For clusters not listed in these papers (probably low mass 
structures), we used T=1.5 keV.

  \section{Global properties  of the various classes }
  
We will consider from this section to the end of the paper only clusters
successfully identified.  

  \subsection{Rich and poor structures }
  
   For X-ray sources unambiguously identified with optical velocity structures, one
   has  then to address the question: has the C1, C2, C3 classification a
   physical basis, or is it only reflecting the X-ray selection process?

\begin{figure}
\centering
\mbox{\psfig{figure=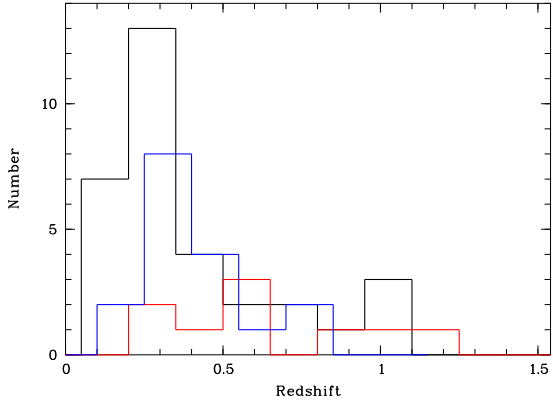,width=8cm,angle=270}}
\caption[]{Redshift distribution for the 3 classes C1 (black
 histogram), C2 (red histogram), C3 (blue histogram).}
\label{fig:fig6}
\end{figure}

   As a first step, we look at the redshift distribution of
   the cluster C1, C2, and C3 classes (Fig.~\ref{fig:fig6}).  For the 32 C1, the 9 
   C2, and the 17 C3 the mean redshift is 0.41, 0.66, and 0.38.  

   Comparing the C1 and C3 distributions and their almost similar mean
   redshifts and letting aside  for a while the z $\geq$ 0.5 C3 structures, it is
   tempting to consider C1 to be in the most cases "X-ray bright and
   optical  nearby ($z \leq 0.4$) rich systems"  and most of the  C3 as "faint
   and poor" at $z \sim 0.4$  redshift. The more distant C3 clusters would be
   rather distant C1-like and therefore "rich". C2 clusters would be a mix of nearby poor 
   and distant rich clusters.

\begin{figure}
\centering
\mbox{\psfig{figure=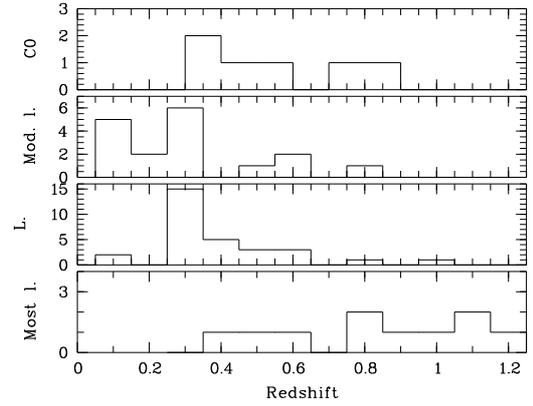,width=8cm,angle=270}}
\caption[]{Redshift distribution for the $most~luminous$ (Most l.), $luminous$ (L.), $moderately~luminous$ 
(Mod. l.), and C0 clusters.}
\label{fig:lclass}
\end{figure}

   We can define alternative categories to the C1, C2, C3 classification. For instance, 
   we chose to group the clusters as a function of their X-ray luminosity. Clusters more luminous than
   10$^{44}$ erg/s were called the X-ray $most~luminous$ sample. Clusters between 10$^{43}$ and 10$^{44}$ erg/s
   were called the X-ray $luminous$ sample. Clusters below 10$^{43}$ erg/s were called the X-ray 
   $moderately~luminous$ 
   sample. Finally, clusters without any X-ray detection (C0 clusters) were considered separately. We give
   in Fig.~\ref{fig:lclass} the redshift distribution of these 4 categories. As expected because of the 
   relatively small angular coverage of the XMM-LSS survey, the $most~luminous$ clusters are mainly
   distant objects. Similarly, $moderately~luminous$ clusters are quite nearby objects because our X-ray
   selection function does not allow us to detect them when they are distant, according to the well known
   Malmquist bias.

\begin{figure}
\centering
\mbox{\psfig{figure=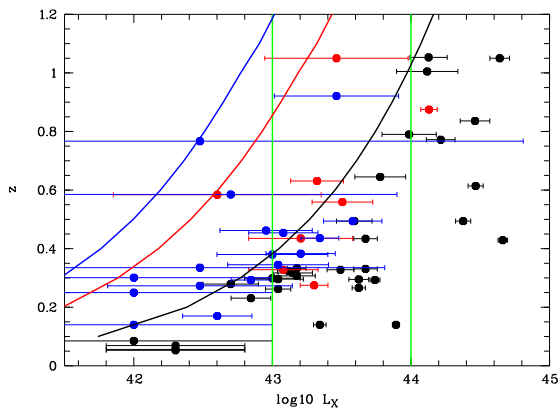,width=8cm,angle=270}}
\caption[]{Present paper cluster distribution in a log10(L$_{X}$) versus redshift diagram. The two vertical green lines
separate the $most~luminous$, $luminous$, and $moderately~luminous$ clusters. Black disks are C1 clusters, red disks are C2
clusters, blue disks are C3 clusters. We also show as black, red, and blue curves the detection limit of the lowest X-ray flux cluster
in C1, C2, and C3 classes.}
\label{fig:refclass}
\end{figure}

We show in Fig.~\ref{fig:refclass} a synthethic view of the clusters listed in Tables~\ref{tab:cand1C1}, 
~\ref{tab:cand1C2}, and ~\ref{tab:cand1C3} allowing the reader to visualize the different classes (C1, C2, C3, 
$most~luminous$, $luminous$, and $moderately~luminous$) in a redshift versus X-ray luminosity diagram.
   
  \subsection{Optical richness }

   We know (e.g. Edge $\&$ Stewart 1991) that optical and X-ray cluster properties should be
   relatively well correlated.
   It is then necessary to characterize the optical richness ($N_{Rich}$) of our clusters.  
   This is done by taking first the number of galaxies in the region of 500 kpc (radius),
   within the photometric redshift slice zmean $\pm$ 0.04  (1+z) and with
   magnitude less than $m^* + 3$. That number is
   then corrected by the "field contribution" estimated in the same manner
   within 1 Mpc to give the final estimate. 
   This richness value is probably not accurate enough in terms of absolute
   value, but can be used in a relative way when comparing a structure to
   another one. We also note that, given the CFHTLS wide magnitude limit we 
   adopted (i'=23), only z$\leq$0.5 clusters are sampled deeply enough to reach
   $m^* + 3$. We therefore only considered these clusters in order to avoid to have
   biased optical richnesses.

   Fitting a richness-velocity dispersion for all z$\leq$0.5 structures for which both data 
   were available, we get: 
   log$(\sigma)$ = ($0.45 \pm 0.24$)  log$(N_{Rich}$)   + ($1.96 \pm 0.38$).

   This is compatible, within the uncertainties, with the value of Yee and
   Ellingson (2003), for similar kind of data: 

     log$(\sigma)$ =  ($0.55 \pm  0.09$)  log$(B_{cg})$  + ($1.26 \pm 0.30$).

\begin{figure}
\centering
\mbox{\psfig{figure=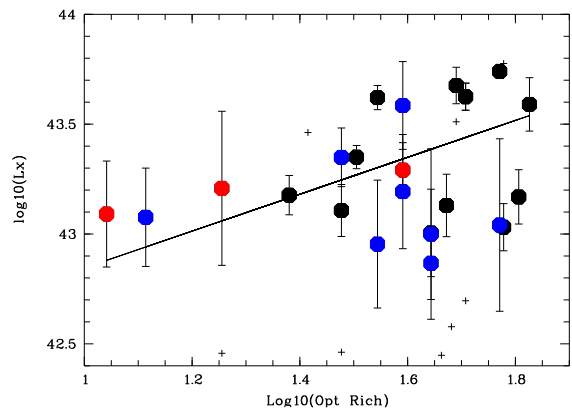,width=8cm,angle=270}}
\caption[]{log($N_{Rich})$ versus log$(L_{X})$. Crosses are clusters
at signal to noise lower than 2 regarding the X-ray luminosity. Disks are
clusters at signal to noise greater than 2 and at z$\leq$0.5 (black: C1, red: C2, blue: C3). 
They give the following fit:  log$(N_{Rich}) = (0.84 \pm 0.51)$   log$(L_{X}) + (7.2 \pm 0.81)$. }
\label{fig:NLx}
\end{figure}

     We now test richness and velocity dispersions versus X-ray
     properties. We first consider z$\leq$0.5 structures with known X-ray luminosity
     and optical richness. We selected only C1, C2 and C3 clusters with X-ray
     luminosity at least two times larger than the associated uncertainty. We show 
     in  Fig.~\ref{fig:NLx} the possible relation between the logarithm
     of $N_{Rich}$ and of Lx. The linear regression between the two parameters has a slope of 
     0.84$\pm$0.51. We note that this value only appears poorly significantly different from a 
     null slope.

     There is a single clear interloper: XLSSC 006 at z$\sim$0.43 (outside of the box shown in 
     Fig.~\ref{fig:NLx}).
     This is one of the most massive clusters in our sample. The observed spectra in the 
     cluster center do not show any sign of AGN activity, so we have no reason to believe 
     that the X-ray flux is polluted by a point source. This cluster shows signs of major 
     substructures in the velocity distribution and this may explain its 
     relatively high Lx value compared to its optical richness. Resulting compression in 
     the intracluster medium could increase the gas density resulting in an enhanced X-ray
     luminosity.

\begin{figure}
\centering
\mbox{\psfig{figure=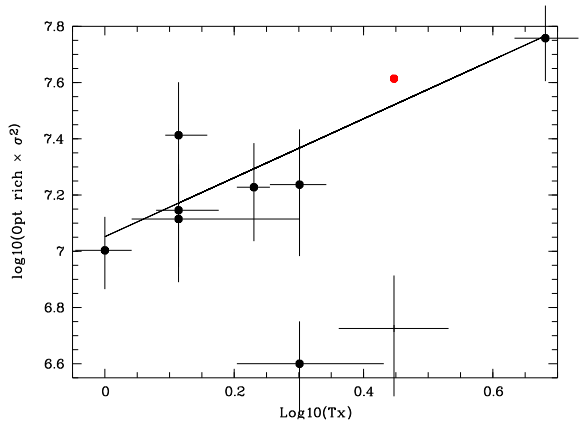,width=8cm,angle=270}}
\caption[]{log$(N_{Rich} \sigma^2)$ versus log$(T_{X})$ for the z$\geq$0.5 clusters (all C1
but XLSSC 046 which is C2). The + sign
  indicates XLSSC 027  and translates to the red disk when replacing the galaxy
velocity dispersion by the weak lensing estimate from Gavazzi $\&$ Soucail (2007). The 
continuous line is the fit (computed without XLSSC 027 and XLSSC 018):
  log$(N_{Rich} \sigma^2) = (1.05 \pm 0.13)$   log$(T_{X}) + (7.05 \pm 0.07)$.}
\label{fig:fig7}
\end{figure}

Considering now clusters at z$\leq$0.5 with a known X-ray temperature (from
Pacaud et al. 2007) and a measured
galaxy velocity dispersion, we searched for a relation between $N_{Rich}$, velocity
dispersion, and X-ray temperature. Fig.~\ref{fig:fig7} shows the relation
between log$(N_{Rich} \sigma^2)$ and log$(T_{X})$. We expect a linear relation as
$(N_{Rich} \sigma^2)$ is at least a qualitative measurement of the kinetic energy of the
clusters, therefore close to the X-ray temperature. Error bars on $(N_{Rich} \sigma^2)$
are 68$\%$ uncertainties and are computed assuming a perfect knowledge of the richness
and the error bars on $\sigma^2$ given in Tables~\ref{tab:cand1C1}, ~\ref{tab:cand1C2}, and 
~\ref{tab:cand1C3}. As quoted in Table~\ref{tab:cand1C1}, these uncertainties are computed 
with a bootstrap technique.

We have two outliers: XLSSC 027 and 
XLSSC 018. XLSSC 027 is known to have strong discrepancies between galaxy and weak lensing 
equivalent velocity dispersions (898$_{-527}^{+523}$km/s from Gavazzi $\&$ Soucail (2007) against 
323$\pm$78 km/s for our own galaxy velocity dispersion and 447$_{-52}^{+82}$km/s for the  Hamana et al. 
(2009) galaxy velocity dispersion). We note that using the weak lensing 
equivalent velocity dispersion puts XLSSC 027 close to the best fit relation. We also
note that this cluster has close contaminants at z=0.31 and 0.38 detected along the line of sight.
This could also affect the measurement of the optical richness via the background estimate.

XLSSC 018 (without any sign of major substructures: see below) would need a larger 
optical richess and/or a larger galaxy velocity dispersion, or a lower X-ray temperature 
to fall on the best fit relation. The last solution in unlikely as only an X-ray temperature 
of the order of 0.4 keV would place XLSSC 018 on the best fit relation. 
A possible explanation would be that we are dealing with a
structure close to a fossil group (even if it does not satisfy the characteristics of
this class of structures). A significant part of the cluster member galaxies could
have merged with the central galaxy, then depopulating the  $\leq$$m^* + 3$ magnitude range
and diminishing the measured optical richness. 

In conclusion, and despite a few detected interlopers, the global agreements show the 
statistical reliability of our optical richness and galaxy velocity dispersion estimates. 

\begin{figure}
\centering
\mbox{\psfig{figure=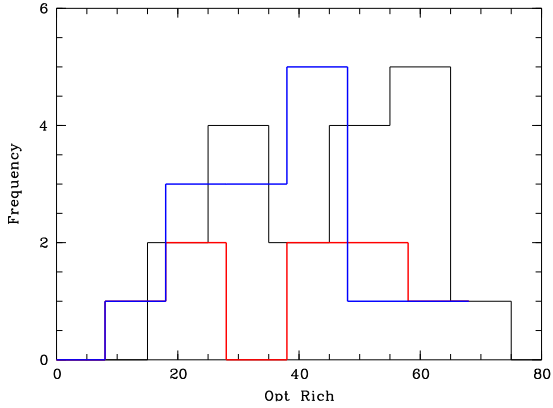,width=8cm,angle=270}}
\caption[]{Distribution of the optical richness for the 3 classes C1 (thin black), C2 (thick red), 
and C3 (thick blue).}
\label{fig:fig8}
\end{figure}

 Fig.~\ref{fig:fig8} shows the histograms of the richness for the 3 classes. C1 has a
  mean $N_{Rich}$  of 45,  C2  a mean $N_{Rich}$ of 37 and C3  a median
  $N_{Rich}$ of 35.

\subsection{Substructure level in velocity space}

Our spectroscopic catalogs are generally too sparse to allow precise
substructure analyses. However, limiting ourselves to the confirmed clusters
with available CFHTLS data and with more than 9 redshifts in the structure 
(10 clusters: XLSSC 013, XLSSC 025, XLSSC 022, XLSSC 006, XLSSC 008, XLSSC 001, XLSSC 000, XLSSC 018,
XLSSC 044, and XLSSC 058), we applied the Serna-Gerbal method (Serna 
$\&$ Gerbal, 1996: SG hereafter) to these spectroscopic catalogs. Two of them
(XLSSC 006 and XLSSC 001) are from the $most~luminous$ cluster category. All the
others (but XLSSC 000 which is a C0 cluster) are members of the $luminous$ cluster 
category. The SG method 
is widely used in order to characterize the substructure level in clusters of 
galaxies (e.g. Adami et al. 2009). Basically, the method allows galaxy subgroups 
to be extracted from a catalog containing positions, magnitudes, and redshift, 
based on the calculation of their relative binding energies. The output of the 
SG method is a list of galaxies belonging to each group, as well as information 
on the binding energy, and mass estimate of the galaxy structures.

The spectroscopic catalogs being still relatively sparse, we will only be able
to detect very prominent substructures, but this is a good way for example to 
check if the analysed clusters are in the process of a major merging event.

Over the 10 analysed clusters, only two (which both belong to the $most~luminous$ cluster 
category) present sign of substructures (XLSSC 006 with two dominant galaxies in its center 
and XLSSC 001) with 2 detected sub-groups. We checked if these two clusters were atypically sampled
in terms of number of available redshifts. XLSSC 001 has 17 redshifts and 
XLSSC 006 16 redshifts. Three other clusters without detected signs of substructures are equally
sampled: XLSSC 013 has 19 redshifts, XLSSC 022 has 15 redshifts, and XLSSC 044 has 17 redshifts.
The substructure detection therefore does not seem to be entirely due to selection effects depending on 
the available number of redshifts. As a conclusion, all
the tested $most~luminous$ clusters show signs of substructures while 
none of the other tested clusters show similar signs.
This would be in good agreement with the most massive clusters being regularly fed by their
surrounding large scale structure in terms of infalling groups. Less luminous clusters
would already be close to their equilibrium, with a less intense infalling activity.
This has, however, to be confirmed with larger spectroscopic samples.

\subsection{Relation between XMM-LSS clusters and their parent cosmic web
  portion}

The previous subsection naturally raises the question of the characteristics of 
cosmological surrounding filaments.
Numerical simulations place clusters of galaxies at the nodes of
the cosmic web. Clusters are then growing via accretion of matter flowing
along the cosmic filaments. This unquestionable scenario for massive clusters
is less evident for low mass structures as groups. Such groups could also form
along the cosmic filaments as for example suggested for fossil groups by Adami
et al. (2007a). Moreover, even for the most massive structures, the precise process of
filament matter accretion is assessed most of the time only by individual cluster
studies (e.g. Bou\'e et al. 2008). 
The XMM-LSS cluster sample presented in this paper offers a unique opportunity
to investigate the cluster-filament connection with a well controled sample.

\subsubsection{General counting method}

We first have to detect the filaments connected to a given cluster. These
filaments are very low mass and young dynamical structures. It is therefore
very difficult to detect them through X-ray observations. This is possible only
in a few peculiar cases (e.g. Bou\'e et al. 2008, Werner et al. 2008) and with 
very long integration times. The
XMM-LSS exposure times are anyway not well suited to such detections. We,
therefore, used optical CFHTLS photometric redshift catalogs.

\begin{landscape}
\begin{table}
\caption{List of the XMM-LSS C1 systems having successfully passed the spectroscopic identification. 
Name refers to the official IAU name. XLSSC refers to the official 
  XMM-LSS name. PH gives the availability of
  CFHTLS T0004 photometric redshifts: 0/1= not available/available. RA and DEC are the decimal J2000 
  coordinates of the X-ray emission center. N is the number
  of galaxies with spectroscopic redshifts belonging to the identified
  structure and within a radius of 500 kpc. ZBWT is the biweight estimate of
  the mean redshift of the structure (at a 0.001 precision).  ERRZ is the upper value of the 
  bootstrap uncertainty on ZBWT, at a precision of 0.001. It was computed only when more than 5 redshifts 
  were available. SIG-GAP is the "Gapper" estimate of the velocity dispersion given
  in km/s. ERR is the bootstrap uncertainty on SIG (computed
  when $\geq$5 redshifts were available). Flux is the value of the [0.5;2] keV flux in 0.5Mpc (radius)
  in 10$^{-14}$ ergs/cm$^2$/s. Lx is the bolometric (0.1 to 50 keV) 
  X-ray luminosity (in 10$^{43}$erg/s) derived from the observed flux. N$_{NP}$ 
  is the net number of photons in 0.5 Mpc. The two last columns
  give name and redshift (when available) when the considered cluster was also detected
  by Hamana et al. (2009), Ueda et al. (2008), or Finoguenov et al. (2010). When 
  coming from  Hamana et al. (2009), the cluster name has the SL Jhhmm.mddmm format.
  When coming from Ueda et al. (2008), the cluster name refers to this paper (4 digits number).
  When coming from Finoguenov et al. (2010), the cluster name refers to this paper (with the SXDF root). 
  The * symbols indicate that the cluster validation was made with one or two spectroscopic
  redshifts. The (l) attached to the cluster
 id means that we have a lack of precision in the measured galaxy redshifts, preventing us
to compute uncertainty of the mean cluster redshift, and velocity dispersions.}
\begin{center}
\begin{tabular}{rlcccccccccccll}
\hline
Name & XLSSC & PH & RA & DEC & N & ZBWT & ERRZ & SIG &
ERR & Flux [0.5;2]keV  & L$_{Bol}$ & N$_{NP}$ & Lit Id & Lit z \\
    &        &      & deg & deg &   &     &      & km/s &
km/s & 10$^{-14}$ ergs/cm$^2$/s & 10$^{43}$erg/s  & &  &  \\
    &        &      &  &  &   &     &      &  &
 & in 0.5Mpc & in 0.5Mpc & in 0.5Mpc  &   &  \\
\hline
XLSSU J021735.2-051325 & 059 & 1 & 34.391 & -5.223 & 8 & 0.645 & 0.001 & 513 & 151  &  1.6$\pm$0.3 & 6.0$\pm$1.1  &  104$\pm$19 & SXDF69XGG/0514 & 0.645 \\
XLSSU J021945.1-045329 & 058 & 1 & 34.938 & -4.891 & 9 & 0.333 & 0.001 & 587 & 236 &  2.0$\pm$0.2 & 1.5$\pm$0.1 & 601$\pm$54  & SXDF36XGG/1176  & 0.333 \\
XLSS J022023.5-025027 & 039 & 0 & 35.098 & -2.841 & 4 & 0.231 &     &  &   &  2.2$\pm$0.4 & 0.7$\pm$0.1   &  183$\pm$34 &  & \\
XLSS J022045.4-032558 & 023 & 0 & 35.189 & -3.433 & 3 & 0.328 &     &  &   &  4.1$\pm$0.4 & 3.1$\pm$0.3   &  465$\pm$43 &  & \\
XLSS J022145.2-034617 & 006 & 1 & 35.438 & -3.772 & 16 & 0.429 & 0.001 & 977 & 157 &  24.1$\pm$0.8 & 46.0$\pm$1.6 &  1409$\pm$52 &  & \\
XLSS J022205.5-043247 & 040* & 1 & 35.523 & -4.546 & 2 & 0.317 &     &  &   &  2.0$\pm$0.3 & 1.4$\pm$0.2   &  265$\pm$35 &  & \\
XLSS J022206.7-030314 & 036* & 0 & 35.527 & -3.054 & 2 & 0.494 &     &  &    &  10.2$\pm$0.5 & 23.8$\pm$1.3  &  551$\pm$32 &  & \\
XLSS J022210.7-024048 & 047 & 0 & 35.544 & -2.680 & 14 & 0.790 & 0.001 & 765 & 163 &  1.4$\pm$0.3 & 9.7$\pm$1.9  &  92$\pm$18 &  & \\
XLSS J022253.6-032828 & 048* & 0 & 35.722 & -3.474 & 2 & 1.005 &     &  &   &  1.1$\pm$0.2 & 13.1$\pm$2.9   &  81$\pm$18 &  & \\
XLSS J022348.1-025131 & 035* & 0 & 35.950 & -2.858 & 1 &  0.069   &     &     & &  6.1$\pm$0.6 & 0.2$\pm$0.1  &  478$\pm$48 &  & \\
XLSS J022356.5-030558 & 028 & 0 & 35.985 & -3.100 & 8 &   0.296  &  0.001   &  281 & 46 &  2.0$\pm$0.4 & 1.1$\pm$0.2     &  177$\pm$31 &  & \\
XLSS J022357.4-043517 & 049* & 1 & 35.989 & -4.588 & 1 &  0.494   &     &     & &  1.9$\pm$0.2 & 3.9$\pm$0.5  &  287$\pm$34 &  & \\
XLSS J022402.0-050525 & 018 & 1 & 36.008 & -5.090 & 9 & 0.324 & 0.001 & 364 & 69  &  1.7$\pm$0.2 & 1.3$\pm$0.1 &  382$\pm$44 &  & \\
XLSS J022404.1-041330 & 029(l) & 1 & 36.017 & -4.225 & 5 & 1.050 &  &  &  &  3.1$\pm$0.2 & 43.7$\pm$3.1 &  323$\pm$24 &  & \\
XLSS J022433.8-041405 & 044 & 1 & 36.141 & -4.234 & 17 & 0.262 & 0.001 & 483 & 100 &  2.6$\pm$0.3 & 1.1$\pm$0.1 &  510$\pm$54 & SL J0224.5-0414 & 0.2627 \\
XLSS J022456.2-050802 & 021 & 1 & 36.234 & -5.134 & 7 & 0.085 & 0.001 & 231 & 64 &  4.2$\pm$0.7 & 0.1$\pm$0.1  &  664$\pm$115 &  & \\
XLSS J022457.1-034856 & 001 & 1 & 36.238 & -3.816 & 17 & 0.614 & 0.001 & 940 & 141  &  7.8$\pm$0.4 & 29.3$\pm$1.6  &  671$\pm$36 &  & \\
XLSS J022520.8-034805 & 008 & 1 & 36.337 & -3.801 & 11 & 0.299 & 0.001 & 544 & 124 &  1.8$\pm$0.4 & 1.0$\pm$0.2  & 196$\pm$39  &  & \\
XLSS J022524.8-044043 & 025 & 1 & 36.353 & -4.679 & 10 & 0.266 & 0.001 & 702 & 178 &  8.7$\pm$0.5 & 4.2$\pm$0.2 &  1098$\pm$62 & SL J0225.3-0441 & 0.2642 \\
XLSS J022530.6-041420 & 041 & 1 & 36.377 & -4.239 & 6 & 0.140 & 0.002 & 899 & 218 &  21.8$\pm$1.1 & 2.2$\pm$0.1 &  1143$\pm$62 & SL J0225.4-0414 & 0.1415 \\
XLSS J022532.2-035511 & 002 & 1 & 36.384 & -3.920 & 8 & 0.772 & 0.001 & 296 & 56 &  2.6$\pm$0.3 & 16.4$\pm$1.7  &  238$\pm$25 &  & \\
XLSSU J022540.7-031123 & 050 & 0 & 36.419 & -3.189 & 13 & 0.140 & 0.001 & 408 & 96 &  54.0$\pm$1.1 & 7.8$\pm$0.2 &  4929$\pm$103 & SL J0225.7-0312 & 0.1395 \\
XLSS J022559.5-024935 & 051 & 0 & 36.498 & -2.826 & 6 & 0.279 & 0.001 & 369 & 99 &  1.1$\pm$0.3 & 0.5$\pm$0.1 &  160$\pm$41 &  & \\
XLSS J022609.9-045805 & 011 & 1 & 36.541 & -4.968 & 8 & 0.053 & 0.001 & 83 & 16 &  13.1$\pm$1.5 & 0.2$\pm$0.1  &  1706$\pm$192 &  & \\ 
XLSS J022616.3-023957 & 052 & 0 & 36.568 & -2.665 & 5 & 0.056 & 0.001 & 194  & 60  &  16.4$\pm$1.8 & 0.2$\pm$0.1   &  1297$\pm$146 &  & \\
XLSS J022709.2-041800 & 005* & 1 & 36.788 & -4.300 & 2 & 1.053 &     &  &    &  1.0$\pm$0.1 & 13.4$\pm$1.8  &  165$\pm$23  & & \\
XLSS J022722.4-032144 & 010 & 0 & 36.843 & -3.362 & 5 & 0.331 & 0.001 & 315 & 56 & 5.7$\pm$0.4 & 4.7$\pm$0.4 &  459$\pm$38 &  & \\
XLSS J022726.0-043216 & 013 & 1 & 36.858 & -4.538 & 19 & 0.307 & 0.001 & 397 & 58 &  2.6$\pm$0.3 & 1.5$\pm$0.2  &  374$\pm$48 &  &  \\
XLSS J022738.3-031758 & 003 & 0 & 36.909 & -3.299 & 9 & 0.836 & 0.001 & 784 & 189 &  3.7$\pm$0.4 & 29.0$\pm$3.1  &  196$\pm$22 &  & \\
XLSS J022739.9-045127 & 022 & 1 & 36.916 & -4.857 & 15 & 0.293 & 0.001 & 535 & 106 &  9.6$\pm$0.3 & 5.5$\pm$0.2 & 1741$\pm$58  &  &  \\
XLSS J022803.4-045103 & 027 & 1 & 37.014 & -4.851 & 6 & 0.295 & 0.001 & 323 & 78 &  6.3$\pm$0.4 & 4.2$\pm$0.3 &  653$\pm$41 & SL J0228.1-0450 & 0.2948 \\
XLSS J022827.0-042547 / & 012 & 1 & 37.114 & -4.432 & 5 & 0.434 & 0.002 & 726 & 95 &  3.4$\pm$0.3 & 4.7$\pm$0.4  &  444$\pm$37 &  & \\
XLSS J022827.8-042601 &  &  &  &  &  &  &  &  &  &  &  & \\
\hline
\end{tabular}
\end{center}
\label{tab:cand1C1}
\end{table}
\end{landscape}

\begin{landscape}
\begin{table}
\caption{Same as Table~\ref{tab:cand1C1} but for C2 XMM-LSS systems. XLSS J022756.3-043119 would 
require more spectroscopy for confirmation. XLSSC 009, 064, and
065 were originally classified as C2, but would be classified as C1 using more recent pipeline version.
The * symbols indicate that the cluster validation was made with one or two spectroscopic
  redshifts. The (l) attached to the cluster
 id means that we have a lack of precision in the measured galaxy redshifts, preventing us
to compute uncertainty of the mean cluster redshift, and velocity dispersions.}
\begin{center}
\begin{tabular}{rlcccccccccccll}
\hline
Name & XLSSC & PH & RA & DEC & N & ZBWT & ERRZ & SIG &
ERR & Flux [0.5;2]keV  & L$_{Bol}$ & N$_{NP}$ & Lit Id & Lit z \\
    &        &      & deg & deg &   &     &      & km/s &
km/s & 10$^{-14}$ ergs/cm$^2$/s & 10$^{43}$erg/s  & &  &  \\
    &        &      &  &  &   &     &      &  &
 & in 0.5Mpc & in 0.5Mpc & in 0.5Mpc  &   &  \\
\hline
XLSSU J021658.9-044904 & 065    & 1 & 34.245 & -4.821 & 3 & 0.435 &     &  &   &   1.1$\pm$0.4 & 1.6$\pm$0.6 &  50$\pm$18 &285/287 & \\
XLSSU J021832.0-050105 & 064   & 1 & 34.633 & -5.016 & 3 & 0.875 &     &  &   &  1.6$\pm$0.1 & 13.5$\pm$0.8 &  937$\pm$56 & SXDF46XGG/829  & 0.875 \\
XLSSU J021837.0-054028 & 063*   & 1 & 34.654 & -5.675 & 2 & 0.275 &     &  &   &  4.1$\pm$0.5 & 2.0$\pm$0.2  &  255$\pm$32 &  &\\
XLSS J022303.3-043621 & 046 & 1 & 35.764 & -4.606 & 8 & 1.213 & 0.001 & 595 & 121 &  0.7$\pm$0.1 & 14.0$\pm$3.2  &  82$\pm$19 &  &\\
XLSS J022307.2-041259 & 030 & 0 & 35.780 & -4.216 & 5 & 0.631 & 0.001 & 520 & 158 &  0.6$\pm$0.1 & 2.1$\pm$0.4  &  129$\pm$27 &  &\\
XLSS J022405.9-035512 & 007 & 1 & 36.024 & -3.920 & 5 & 0.559 & 0.001 & 369 & 179 &  1.2$\pm$0.3 & 3.2$\pm$0.7  &  113$\pm$25 &  &\\
XLSSU J022644.2-034107 & 009 & 1 & 36.686 & -3.684 & 8 & 0.328 & 0.001 & 261 & 53 &  1.7$\pm$0.4 & 1.2$\pm$ 0.3 &  87$\pm$21  &  &\\
XLSS J022725.1-041127 & 038 & 1 & 36.854 & -4.191 & 4 & 0.584 &  &  & & 0.1$\pm$0.1 & 0.4$\pm$0.3  &  31$\pm$28  &  &\\
XLSS J022756.3-043119* & - & 1 & 36.985 & -4.522 & 2 & 1.050? &     &  &   &   0.2$\pm$0.1 & 2.9$\pm$1.5  &  49$\pm$26  &  &\\
\hline
\end{tabular}
\end{center}
\label{tab:cand1C2}
\end{table}
\end{landscape}

\begin{landscape}
\begin{table}
\caption{Same as Table~\ref{tab:cand1C2} but for C3 XMM-LSS systems. The ** symbols 
indicate that we merged by hand 2 groups separated by the gapper technique in the final
analysis. The last line (C555 class) is the new cluster (see 
text). The * symbols indicate that the cluster validation was made with one or two spectroscopic
  redshifts. The (l) attached to the cluster
 id means that we have a lack of precision in the measured galaxy redshifts, preventing us
to compute uncertainty of the mean cluster redshift, and velocity dispersions.}
\begin{center}
\begin{tabular}{rlcccccccccccll}
\hline
Name & XLSSC & PH & RA & DEC & N & ZBWT & ERRZ & SIG &
ERR & Flux [0.5;2]keV  & L$_{Bol}$ & N$_{NP}$ & Lit Id & Lit z \\
    &        &      & deg & deg &   &     &      & km/s &
km/s & 10$^{-14}$ ergs/cm$^2$/s & 10$^{43}$erg/s  & &  &  \\
    &        &      &  &  &   &     &      &  &
 & in 0.5Mpc & in 0.5Mpc & in 0.5Mpc  &   &  \\
\hline
XLSSU J021651.3-042328*  & - & 1 & 34.214 & -4.392 & 1 &  0.273   &     &     & &  0.6$\pm$0.3 & 0.3$\pm$0.2  &  58$\pm$32  &  &\\
XLSSU J021754.6-052655(l)  & 066 & 1 & 34.478 & -5.447 & 6 & 0.250 &        &  & &  0.3$\pm$0.3 & 0.1$\pm$0.1 &  68$\pm$59 & SXDF85XGG/621 & 0.25 \\
XLSSU J021842.8-053254  & 067 & 1 & 34.678 & -5.548 & 5 & 0.380 &  0.001   & 847 & 279 &  1.0$\pm$0.4 & 1.0$\pm$0.4  &  84$\pm$34 & SXDF01XGG/876  & 0.378 \\
XLSSU J021940.3-045103*  & - & 1 & 34.919 & -4.852 & 1 &   0.454     &        &    &  &   0.8$\pm$0.2 & 1.2$\pm$0.3  &  193$\pm$44 &  &\\
XLSS J022258.4-04070  & 024 & 1 & 35.744 & -4.121 & 5 & 0.293 & 0.001 & 452 & 98 &  1.3$\pm$0.2 & 0.7$\pm$0.1  &  254$\pm$41  &  &\\
XLSS J022341.8-043051 & 026 & 1 & 35.925 & -4.514 & 3 & 0.436 &     &  &  &  1.6$\pm$0.2 & 2.2$\pm$0.3  &  255$\pm$34  &  &\\
XLSSU J022509.2-043239 & 037 & 1 & 36.286 & -4.542 & 3 & 0.767 &     &  &  &  0.1$\pm$0.1 & 0.3$\pm$0.7  &  11$\pm$28  &  &\\
XLSSU J022510.5-040147 & 043** & 1 & 36.294 & -4.029 & 3 & 0.170 &     &  &  &  2.6$\pm$0.9 & 0.4$\pm$0.1  &  142$\pm$49  &  &\\
XLSS J022522.8-042649 & 042 & 1 & 36.345 & -4.447 & 6 & 0.462 & 0.003 & 1009 & 257 &  0.6$\pm$0.2 & 0.9$\pm$0.3  &  111$\pm$33 &  &\\
XLSS J022542.2-042434  & 068 & 1 & 36.424 & -4.410 & 4 & 0.585 &     &  & &  0.2$\pm$0.2 & 0.5$\pm$0.6   &  23$\pm$28 &  &\\
XLSS J022610.0-043120  & 069 & 1 & 36.542 & -4.523 & 8 & 0.824 & 0.001 & 398 & 114 & -0.4$\pm$0.2 & -  & - &  &\\
XLSSU J022627.3-050001 & 017 & 1 & 36.615 & -5.003 & 5 & 0.383 & 0.001 & 352 & 132 &  1.5$\pm$0.4 & 1.6$\pm$0.4  &  134$\pm$35 &  &\\
XLSSU J022632.5-040314 & 014 & 1 & 36.635 & -4.054 & 6 & 0.345 & 0.001 & 304 & 84 &  1.4$\pm$0.5 & 1.1$\pm$0.4  &  74$\pm$29  &  &\\
XLSSU J022632.4-050003 & 020 & 1 & 36.638 & -5.007 & 3 & 0.494 &     &  &  &  2.0$\pm$0.4 & 3.8$\pm$0.8  &  159$\pm$32  &  &\\
XLSSU J022726.8-045412  & 070 & 1 & 36.860 & -4.904 & 5 & 0.301 & 0.001 & 180 & 47 &  0.2$\pm$0.2 & 0.1$\pm$0.1  &  32$\pm$39 &  &\\
XLSS J022754.1-035100*  & - & 1 & 36.974 & -3.851 & 2 & 0.140 &     &  &  & 0.7$\pm$0.5 & 0.1$\pm$0.1  &  107$\pm$72  &  &\\
XLSSU J022828.9-045939 & 016* & 1 & 37.121 & -4.994 & 2 & 0.335 &     &  & &  0.4$\pm$0.5 & 0.3$\pm$0.4    &  33$\pm$44 &  &\\
\hline
C555   & - & 1 & 36.375 & -4.429 & 7 & 0.921 & 0.001 & 759 & 248 &  0.3$\pm$0.1 & 2.9$\pm$1.1  &  47$\pm$18  & &\\
\hline
\end{tabular}
\end{center}
\label{tab:cand1C3}
\end{table}
\end{landscape}

\begin{table*}
\caption{Same as Table~\ref{tab:cand1C2} but for C0 clusters. An approximate
upper limit for the X-ray luminosity of these clusters would be the faintest 
detected value for C3 clusters: $\sim$0.08 10$^{43}$ erg/s.}
\begin{center}
\begin{tabular}{cccccccccc}
\hline
Name & XLSSC & PH & RA & DEC & N & ZBWT & ERRZ & SIG &
ERR \\
    &        &      & deg & deg &   &     &      & km/s &
km/s \\
\hline
022207.9-042808*  & - & 1 & 35.533 & -4.469 & 2 & 0.316 &     &  &    \\ 
022402.4-051753 & 000 & 1 & 36.010 & -5.298 & 11 & 0.496 & 0.001 & 435 & 88 \\ 
022405.0-041612  & - & 1 & 36.021 & -4.270 & 8 & 0.862 & 0.001 & 457 & 70 \\ 
022528.3-041536 & 045 & 1 & 36.369 & -4.261 & 4 & 0.556 & & & \\
022550.4-044500*  & - & 1 & 36.460 & -4.750 & 2 &  1.529   &     &     & \\
022647.5-041428*  & - & 1 & 36.698 & -4.241 & 1 &  0.742   &     &     & \\ 
022829.7-031257*  & - & 0 & 37.124 & -3.216 & 2 & 0.313 &     &  &    \\ 
\hline
\end{tabular}
\end{center}
\label{tab:cand1C0}
\end{table*}

  - We first selected
only clusters at z$\leq$0.5 in order to be able to sample deeply enough the
galaxy population to potentially detect the filaments, given the i'=23 magnitude 
limit for the photometric redshift catalog as demonstrated in Adami et al. (2010).

  - For a given cluster, we selected galaxies with photometric redshifts 
in a 0.04$\times$(1+z) slice around the cluster redshift.

  - We then computed the number of galaxies in the slice in 72 angular sectors 
 10 degrees wide each, with position angles between 0 and 360 degrees. Each sector
 was overlapping the previous one by 5 degrees. We did this exercise for
 galaxies in a circle of 2.5Mpc radius, and in an annulus between 2.5 and 5
 Mpc.

\subsubsection{Filament detection and signal enhancement}

Intuitively, if a given sector is significantly more populated than 
other sectors, it means that this sector is including a galaxy overdensity
which could be explained by a filament or by a group along a filament. The
question is then to define a significance level. For a given cluster (and then a
given redshift slice) and a given
radius, we chose to compute the mean and dispersion of the galaxy
numbers in the 72 considered sectors. If a given sector had a number of
galaxies larger than the mean + 3 times the dispersion, we considered this
sector as hosting a potential cosmic filament portion.

However, individual clusters exhibit at best a single 3-$\sigma$ significant
candidate filament. This is due to the intrinsic very low galaxy density in filaments.
Moreover, the goal of the present section is not to make individual
cluster studies, but to draw statistical tendencies. In order to enhance 
the significance of the filament detections, we therefore stacked different 
clusters, considering two categories: the $luminous$ and the $moderately~luminous$ clusters. Other 
categories did not have enough cluster members in the selected redshift range. This technique
is based on the assumption that the angular separation between different 
filaments feeding a given cluster is more or less constant. In order to make
such a stack we now need to homogeneize the cluster position angles. 
We chose the position angle of the highest galaxy overdensity (the PA), limiting ourselves 
to clusters exhibiting a detection more significant than the 3-$\sigma$ level. Selected 
clusters were rotated to have their most significant filament at an arbitrary position 
angle of 180 degrees (east-west direction). In order to check if this alignment technique has a physical
meaning, we superposed in the same way the X-ray images using the
position angles defined by the highest galaxy overdensities (more significant than 
the 3-$\sigma$ level). After rotating these X-ray images, we spatially rescaled them to 
physical units (kpc) according to redshift, and we simply added them together, taking into 
account the corresponding weight maps. The resulting point spread function is a mean of the 
individual values and remains small compared to cluster typical sizes. Fig.~\ref{fig:alignX} 
shows that we generate in this way a clearly elongated synthethic X-ray cluster along the 
180deg direction. The measured ellipticity of the external isophote is equal to 0.41. 
If instead of a simple sum we compute the median of the images (Fig.~\ref{fig:alignX}),
the resulting ellipticity of the external isophote is still 0.36. Finally, if we combine the 
X-ray images without correcting by the optically determined orientation, we produce 
Fig.~\ref{fig:alignXrandom}, showing a basically null ellipticity.

If the galaxy-defined prefered orientations are valid, the detected elongation in X-rays is 
an expected behaviour as X-ray emiting groups are also expected to fall onto clusters coming from
surrounding filaments (see e.g. Bou\'e et al. 2008).

\begin{figure}
\centering
\caption[]{Stacked X-ray images with position angles defined by the 
highest galaxy overdensities aligned along the 180deg arbitary angle. Images were rescaled to 
physical units according to cluster redshift. Image size is 1 Mpc$\times$650 kpc. Contours were
drawn with a 20$\times$20 pixel smoothing. Upper figure: mean stacking. Lower figure: 
median stacking.}
\label{fig:alignX}
\end{figure}

\begin{figure}
\centering
\caption[]{Same as Fig.~\ref{fig:alignX} with a median stacking and without any position angle correction.}
\label{fig:alignXrandom}
\end{figure}

We have to take into account the cluster redshift before merging their
galaxy populations. A single catalog magnitude limit would evidently increase
the weight of nearby clusters as compared to more distant ones. We therefore 
limited the galaxy catalogs to i'=23 at z=0.5. The limits were brighter by $D$ 
magnitudes for nearer clusters with $D$ being the distance moduli
difference between the cluster redshifts and z=0.5. 

Renormalizing finally the galaxy counts by the number of selected clusters in
a given class, we 
are able to produce figures giving the mean galaxy counts as a function of the
angular position.

\subsubsection{Results}

\begin{figure}
\centering
\mbox{\psfig{figure=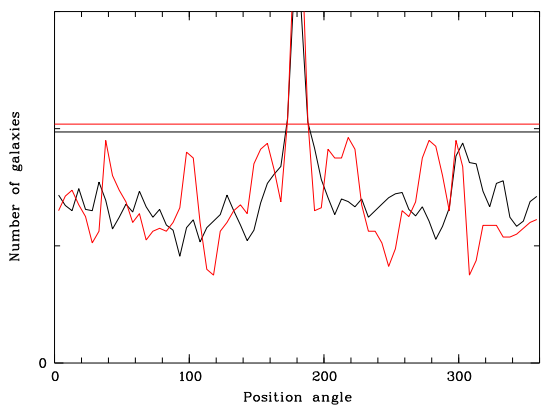,width=8cm,angle=270}}
\caption[]{Stacked angular galaxy counts (in arbitray units) for $luminous$ (black line) and 
$moderately~luminous$ (red line) clusters in annuli of [2.5,5]Mpc. The horizontal lines
show the 3-$\sigma$ detection levels.}
\label{fig:LSS1}
\end{figure}

We first draw stacked (using the previously defined PA of each cluster)
angular galaxy counts for $luminous$ and $moderately~luminous$ 
clusters in the annulus [2.5,5]Mpc (Fig.~\ref{fig:LSS1}).
The minimal and maximal radii have been choosen to be close to the mean
virial radius of clusters (e.g. Carlberg et al. 1996) and not too large in
order to limit the contamination by other clusters. These annuli will
therefore mainly sample the infalling galaxy layers, just before the cluster 
dominated areas. As expected, the signal from the most significant filament candidate
is drastically increased, but no other features are detected at the 3-$\sigma$ level
besides the main filament.

\begin{figure}
\centering
\mbox{\psfig{figure=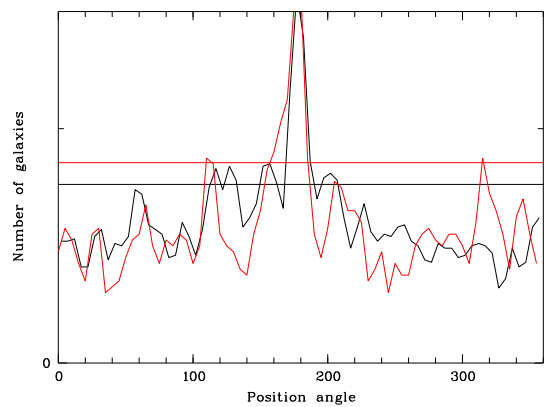,width=8cm,angle=270}}
\caption[]{Stacked angular galaxy counts (in arbitray units) for $luminous$ (black line) and 
$moderately~luminous$ (red line) clusters inside a circle of 2.5 Mpc radius. The 
horizontal lines show the 3-$\sigma$ detection levels.}
\label{fig:LSS2}
\end{figure}

We redo now the same exercise inside a 2.5 Mpc radius central area
(Fig.~\ref{fig:LSS2}). This area is mainly dominated by the clusters themselves (the few
hundreds of kpc central areas) and by
the galaxy layers just beginning to experience the cluster influence (close to
the virial radius). We
therefore investigate the cluster region as fed by the connected filaments. 
The signal from the main filaments is still increased. Other significant filament
candidates are detected at the 3-$\sigma$ level mainly for the $moderately~luminous$
cluster sample.

This difference between the 2.5 Mpc radius central area and the [2.5,5]Mpc annulus could be explained 
if the immediate vicinity of the considered clusters would be depopulated by the potential well of the 
clusters, diminishing the contrast between cosmic filaments and voids. 
Larger spectroscopic redshift samples will soon become available in the area and will 
allow us to refine our results in future works.

\section{Cluster galaxy populations characteristics}
 
We now investigate the optical properties of the galaxy populations in association with the 
X-ray clusters. We refer the reader to Urquhart et al. (2010) for individual studies of 
the clusters providing a temperature measurement.

\subsection{Rest frame Red Sequences}

The so-called red sequence (RS hereafter) commonly shows up to at least z$\sim$1.2 
(e.g. Stanford et
al. 2002) in the massive structures. It is also detected in a less compact state
in field galaxy populations up to z$\sim$2 (Franzetti et al. 2007). We therefore 
searched RSs in our sample of clusters. This sample does not provide 
enough statistics per cluster in order to perform individual studies. The optimal 
strategy is therefore to  build synthetic clusters by gathering galaxies 
for clusters of the same category. We therefore considered 4 classes of clusters:
the $most~luminous$, the $luminous$, the $moderately~luminous$, and the C0 clusters. The RS
being a powerful tool to characterize the evolutionary stage 
of the cluster galaxy populations (e.g. Adami et al. 2007b), such a study will allow
us to assess the properties of these 4 cluster classes. 

In order to be able to stack different clusters at different redshifts,
rest frame absolute magnitudes were computed in the process of getting photometric
redshifts with LePhare (e.g. Ilbert et al. 2006) and we used these magnitudes to 
compute colors. Basically the method 
consists in selecting the observed band which is the closest of the requested rest 
frame band to compute the magnitude in this band applying correction factors. They are described
in the annex of Ilbert et al. (2005), including for example k-correction. This method is 
the closest of the observations and minimizes our dependence on the assumed spectral 
energy distributions, which could not be exactly the same in clusters and in the field 
(see also annex I of the present paper).

\subsubsection{Red sequence using spectrocopic redshifts}

  In a first step we look only at galaxy members using spectroscopic redshifts 
  rather than photometric ones in order to remove potential interloper galaxies which 
  are non cluster members but close to the cluster redshift. Such galaxies could be interpreted as
  cluster members considering only photometric redshifts because of their limited precision.
  We here consider u*-r' rest frame colours and look at their behaviour versus rest frame r'
  absolute magnitude.

\begin{figure}
\centering
\mbox{\psfig{figure=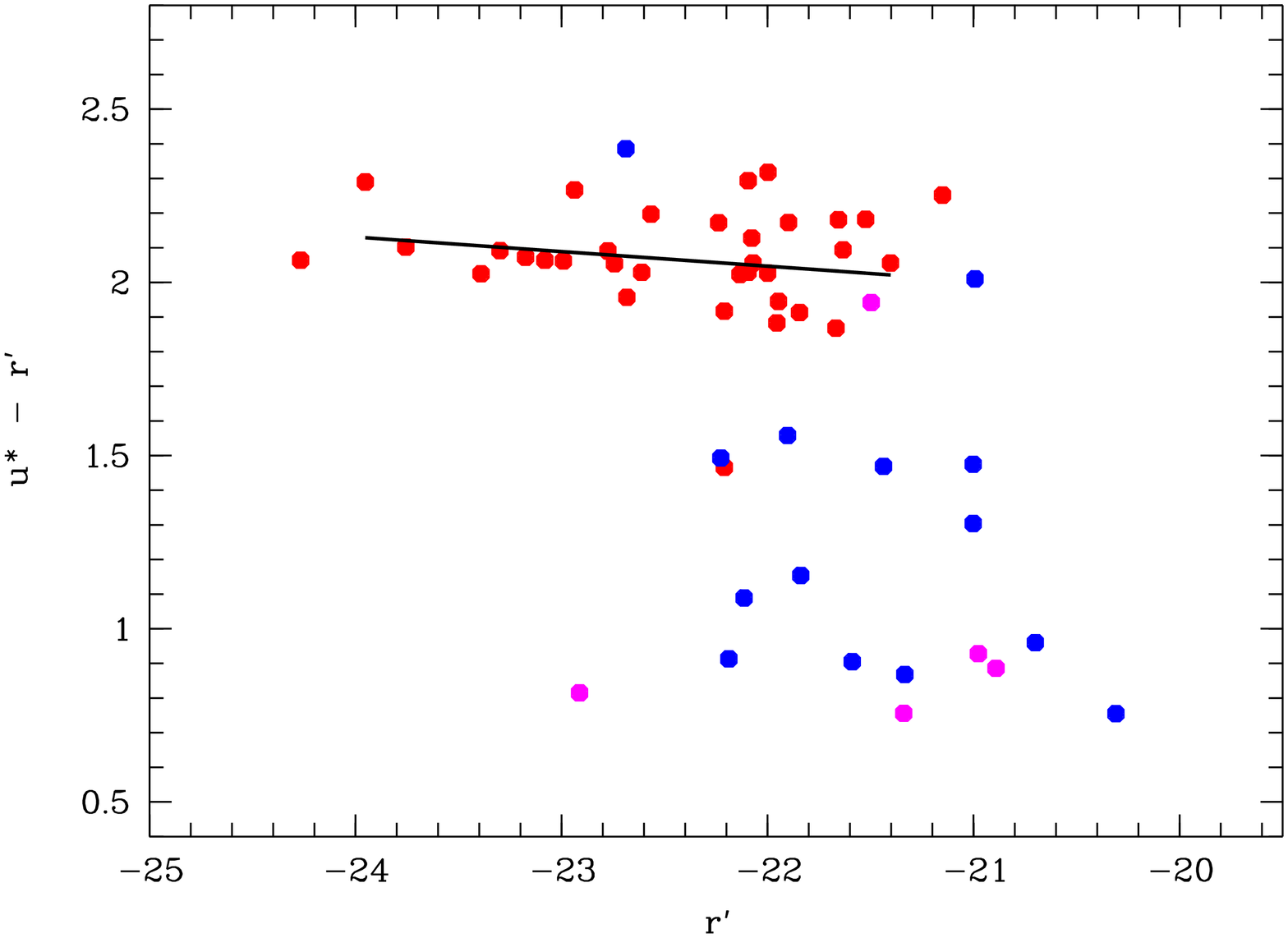,width=7.5cm,angle=270}}
\mbox{\psfig{figure=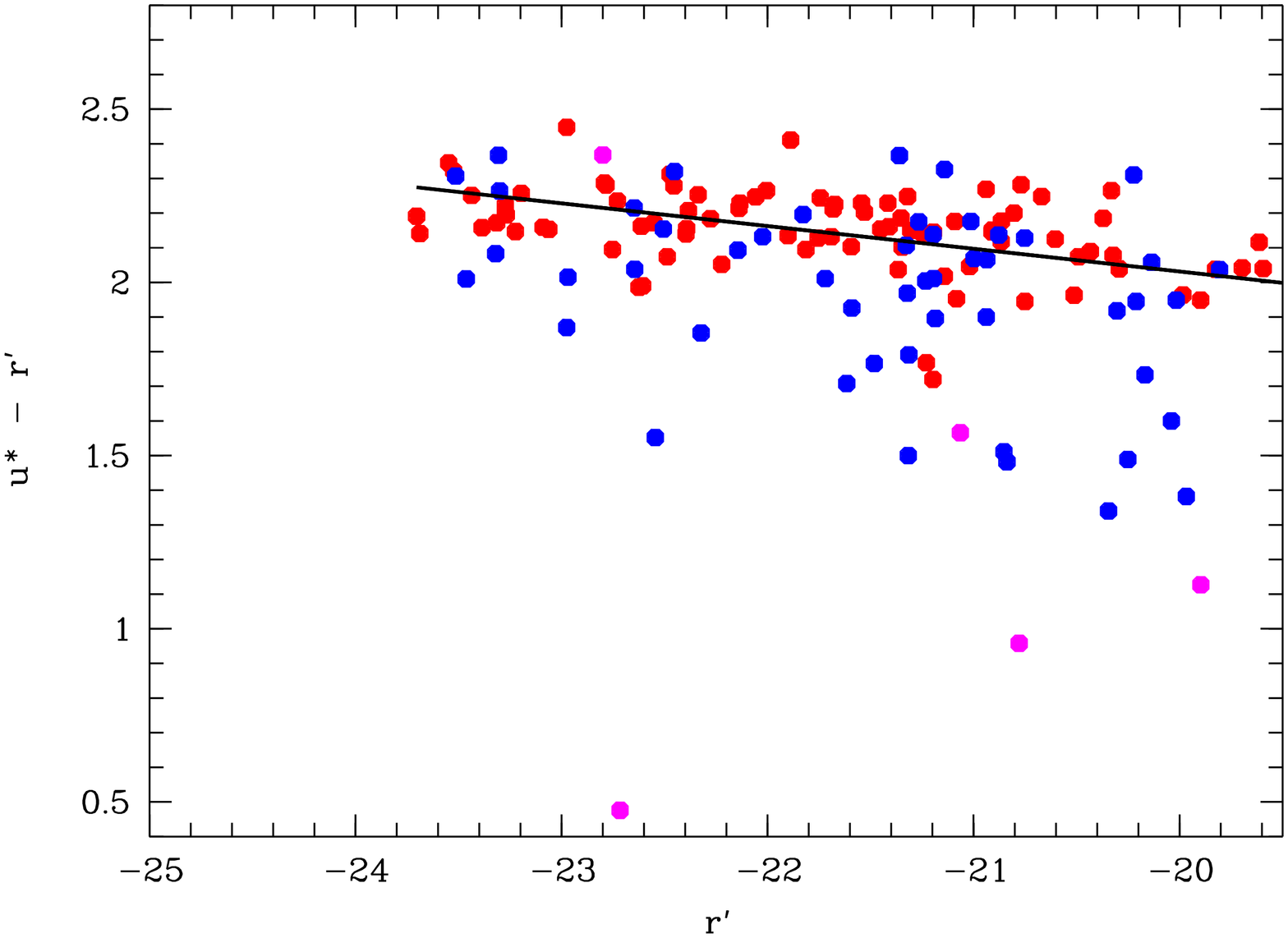,width=7.5cm,angle=270}}
\mbox{\psfig{figure=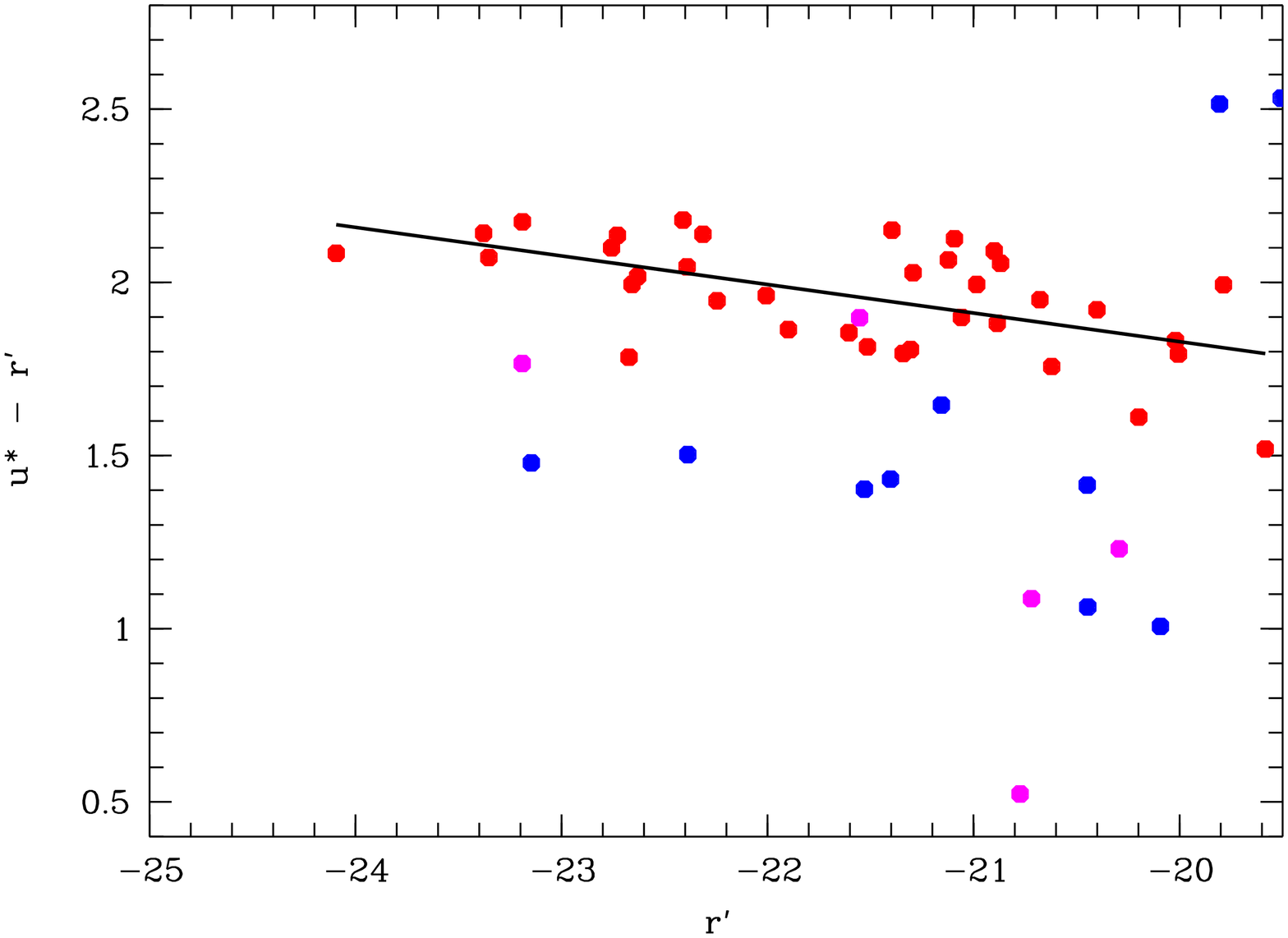,width=7.5cm,angle=270}}
\mbox{\psfig{figure=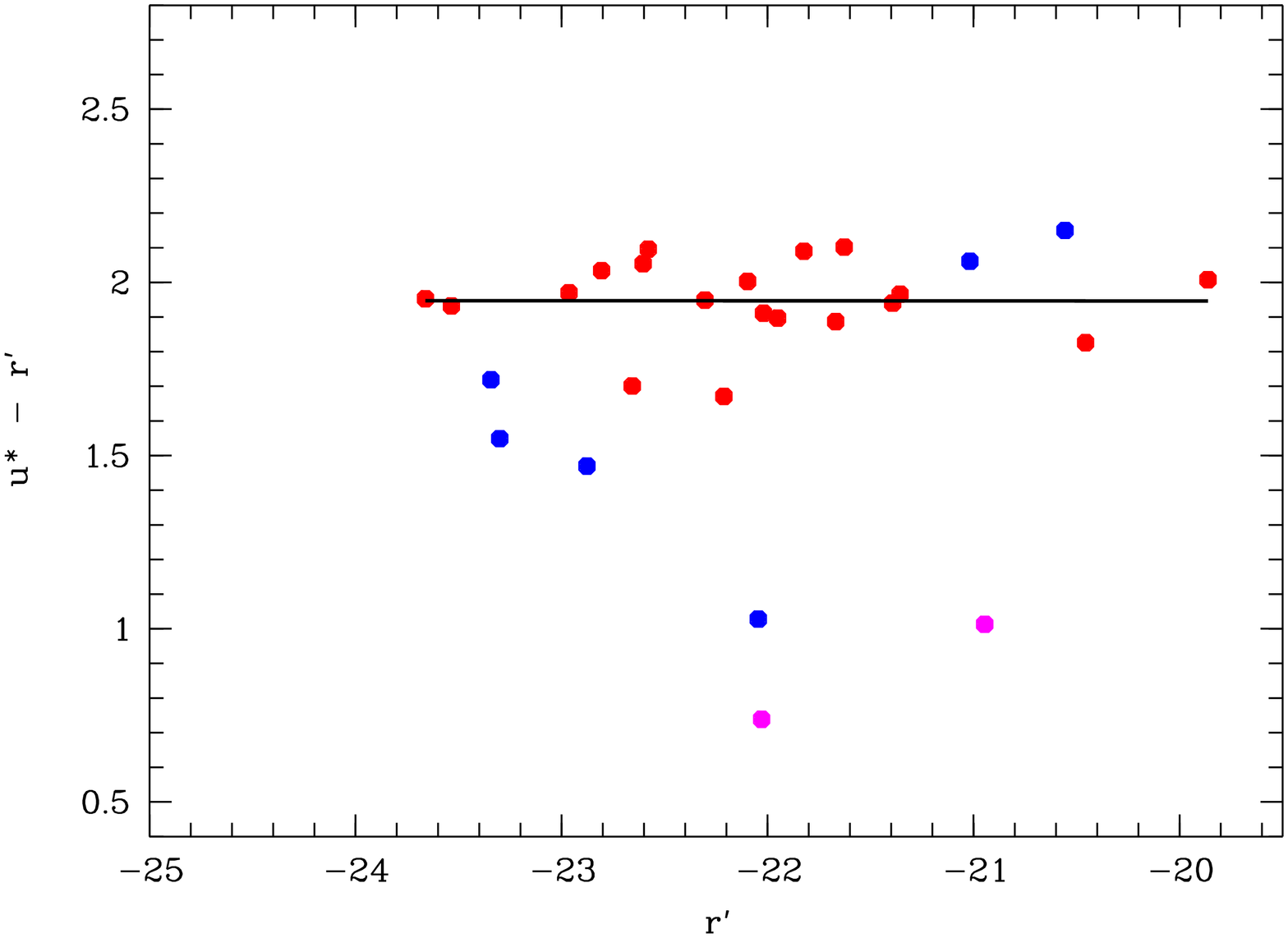,width=7.5cm,angle=270}}
\caption[]{Rest frame u*-r' versus absolute r' magnitude relation for clusters using 
spectroscopic redshifts to compute absolute magnitudes. We only plot cluster members in these figures. 
From top to bottom, figures are for the $most~luminous$,
the $luminous$, the $moderately~luminous$, and the C0 clusters. Red symbols correspond to early type 
galaxies ($T\leq$21), blue symbols correspond to late type non starburst galaxies (58$\geq T \geq$21),
and pink symbols correspond to starburst galaxies (T$\geq$59). Black continuous lines are computed
using only $T\leq$21 galaxies (see Table~\ref{tab:slopecmr}).}
\label{fig:fig9}
\end{figure}     

Fig.~\ref{fig:fig9} shows that a RS is present with u*-r'$\sim$2
for all clusters.  The slopes of the RS appear in good agreement 
with literature estimates (e.g. Adami et al. 2007b: between -0.1 and -0.02 for the 
Coma cluster) and are given in Table~\ref{tab:slopecmr}. As expected, RSs
are populated by early type galaxies, while later type objects are grouped
in a much less compact bluer sequence.

There are potential differences between C0 clusters (without detectable X-ray emission) and other 
classes. The slope of the RS appears nearly flat for C0 clusters while being more negative for more
luminous clusters. This effect is only poorly significant when considering the uncertainty of the slope of 
these RS's. We performed however a bi-dimensional Kolmogorov-Smirnov statistical test on the early type galaxies
of Fig.~\ref{fig:fig9}. The probability that the C0 and the $most~luminous$ cluster early type galaxies come from
the same population is only 0.6$\%$. The probability that the C0 and the $luminous$ cluster early type galaxies come from
the same population is only 0.1$\%$. Finally, the probability that the C0 and the $moderately~luminous$ cluster early type 
galaxies come from the same population is 3.1$\%$. At least for the $most~luminous$ and $luminous$ cluster populations,  
early type galaxies therefore seem to be differently distributed in a color magnitude relation compared to C0
cluster early type galaxies. If these differences come from the slope of the RS, this effect could be interpreted as 
a metallicity effect (Kodama $\&$ Arimoto, 1997). 
The more massive a galaxy, the more easily it will retain metals against dissipative processes. The 
more metals present in a galaxy, the redder the galaxy will be. Massive galaxies are therefore expected 
to be redder than lower mass objects. 
A possible explanation would be that the
faint early type C0 cluster galaxies would originate from depleted cores of larger galaxies, so being metal 
rich before becoming faint (see e.g. Adami et al. 2006). This is possible for example in small groups
where velocity dispersion is low enough to favor galaxy-galaxy encounters.

Galaxie members of the $most~luminous$ clusters also appear to exhibit a more pronounced dichotomy
between early and late type objects. Blue members of the $most~luminous$ clusters are clearly bluer than
blue members of the less luminous clusters.

\begin{table}
\caption{Slopes of the red sequences for the four classes of 
clusters: $most~luminous$, $luminous$,
$moderately~luminous$ and C0. }
\begin{center}
\begin{tabular}{ccccc}
\hline
Category & Slope \\
\hline
$most~luminous$       &  -0.04$\pm$0.04   \\
     $luminous$       &  -0.04$\pm$0.02   \\
$moderately~luminous$ &  -0.10$\pm$0.04   \\
C0                  &  -0.01$\pm$0.05   \\
\hline
\end{tabular}
\end{center}
\label{tab:slopecmr}
\end{table}

\subsubsection{Age of formation of the cluster galaxy stellar populations}

We expect distant clusters to naturally exhibit younger galaxy star populations compared to nearby structures. In 
order to investigate this question, we computed with LePhare ages of stellar population in galaxies with a spectroscopic 
redshift lying inside the considered clusters. The templates used to generate public photometric redshifts in the CFHTLS 
does not allow to provide this information, so we used in LePhare the Bruzual $\&$ Charlot (2003) templates, fixing the 
redshifts to the spectroscopic values. The metallicity was let free to vary between 0.004, 0.008, and 0.02 Z$_{\odot}$. 
In C0 clusters, z=[0.3;0.6] 
galaxies have a stellar population aged of 6.2$\pm$1.9 Gyr, and z=[0.7;0.9] galaxies have a 
stellar population aged of only 2.7$\pm$1.3 Gyr. Considering members of $luminous$ clusters, z=[0.25;0.35] galaxies have a 
stellar population aged of 7.4$\pm$1.0 Gyr, and z=[0.35;0.65] galaxies have a stellar population aged of only 
5.3$\pm$2.1 Gyr. Finally, members of the $most~luminous$ clusters, z=[0.4;0.65] galaxies have a stellar population aged 
of 5.2$\pm$2.1 Gyr, and z=[0.75;1.25] galaxies have a stellar population aged of 3.3$\pm$1.1 Gyr. 

Taking the mean redshift of the highest redshift bin for each
of these 3 categories and diminishing the corresponding elapsed time since the beginning of the Universe by the mean age of
the early type galaxy stellar populations leads us to estimate the mean age of formation of the star populations in these 
galaxies. Galaxy stellar populations probably formed at z$\sim$1.6 in C0 clusters, at z$\sim$2 in $luminous$ clusters, and 
at z$\sim$2.5 in the most $most~luminous$. 
These values are in good agreement with general expectations for the massive clusters to form early than low mass structures,
up to redshifts close to z$\sim$2.

\subsubsection{Red sequence using photometric redshifts and color-color diagrams}

\begin{figure}
\centering
\mbox{\psfig{figure=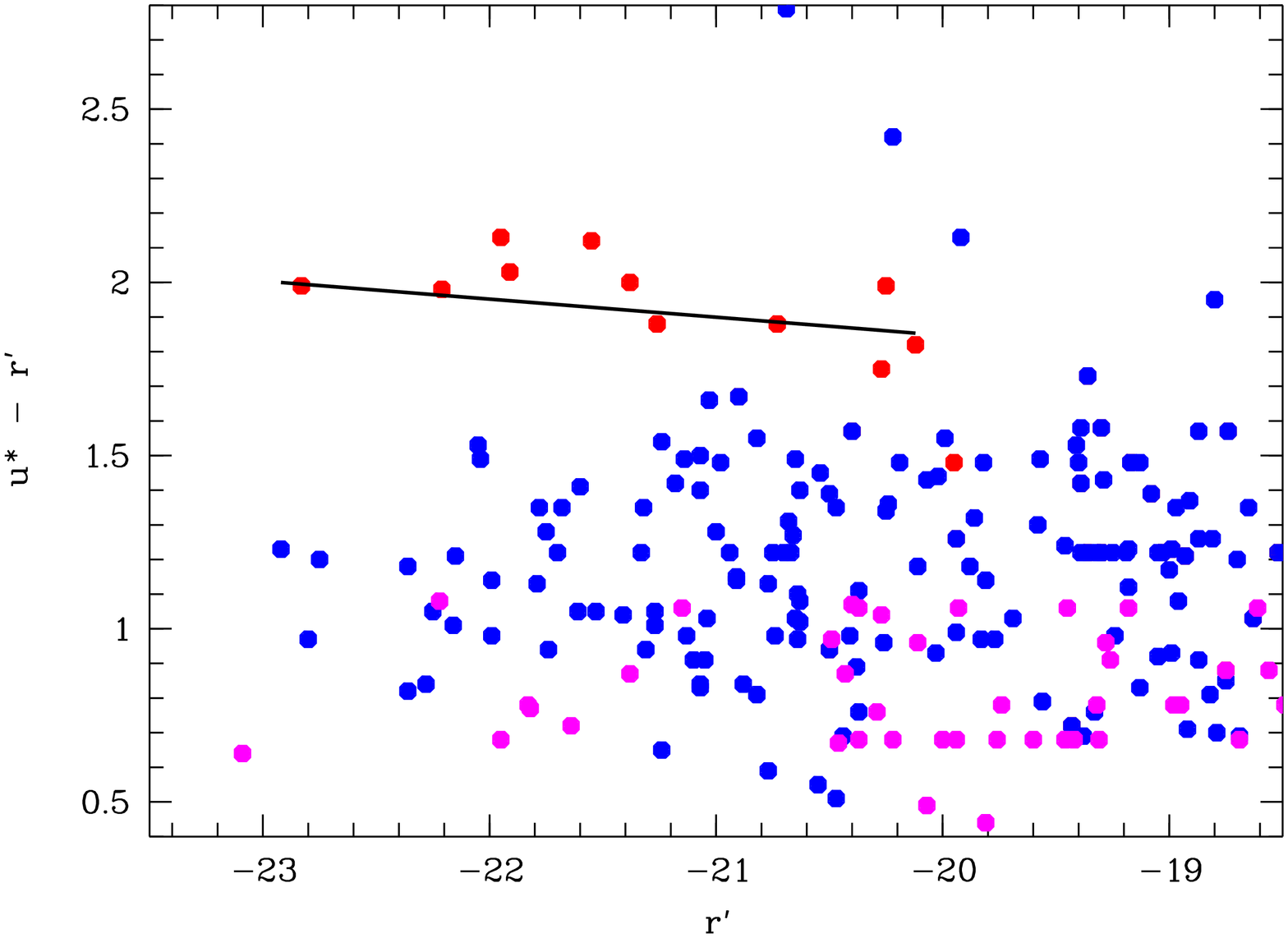,width=8cm,angle=270}}
\mbox{\psfig{figure=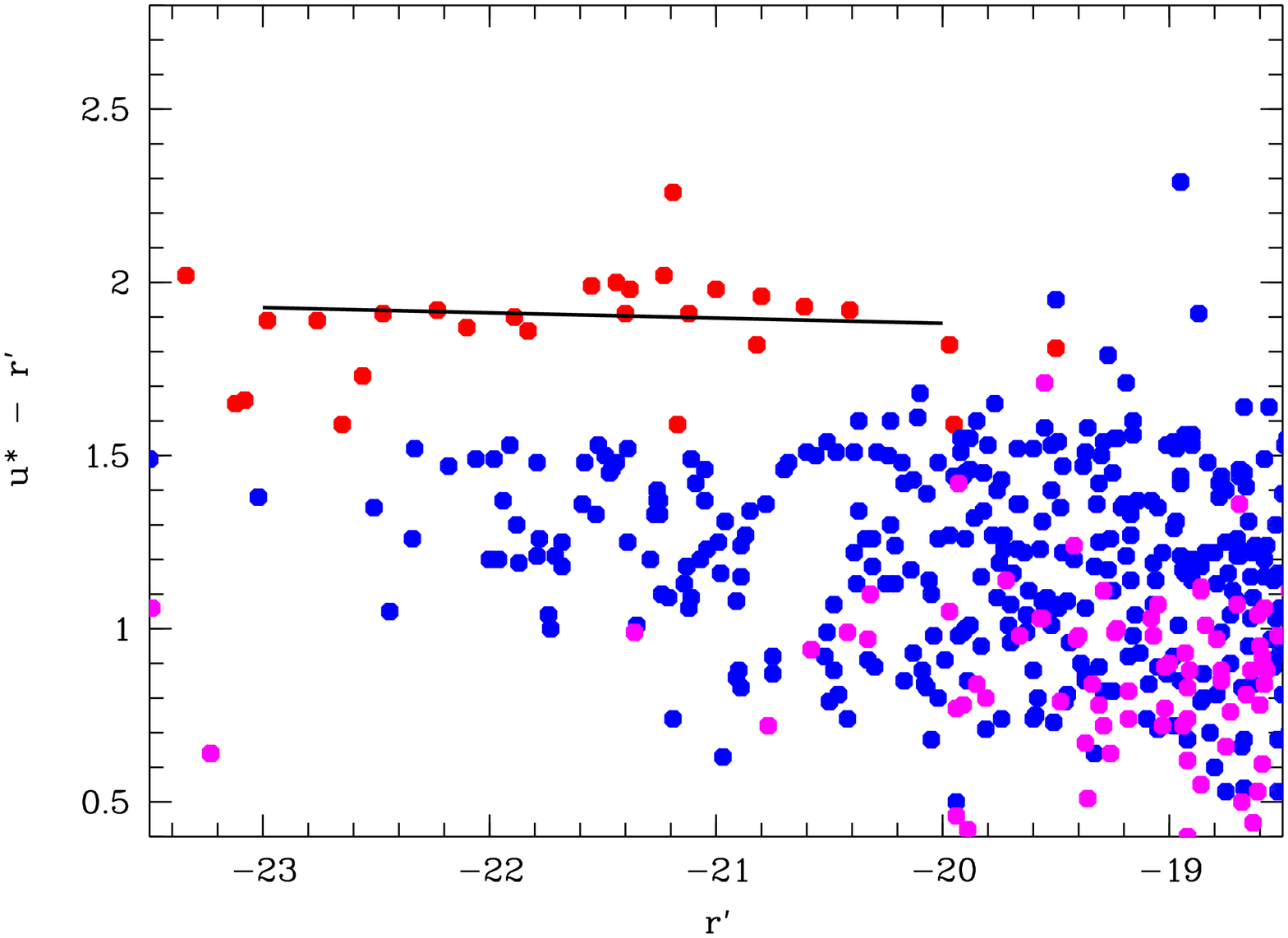,width=8cm,angle=270}}
\mbox{\psfig{figure=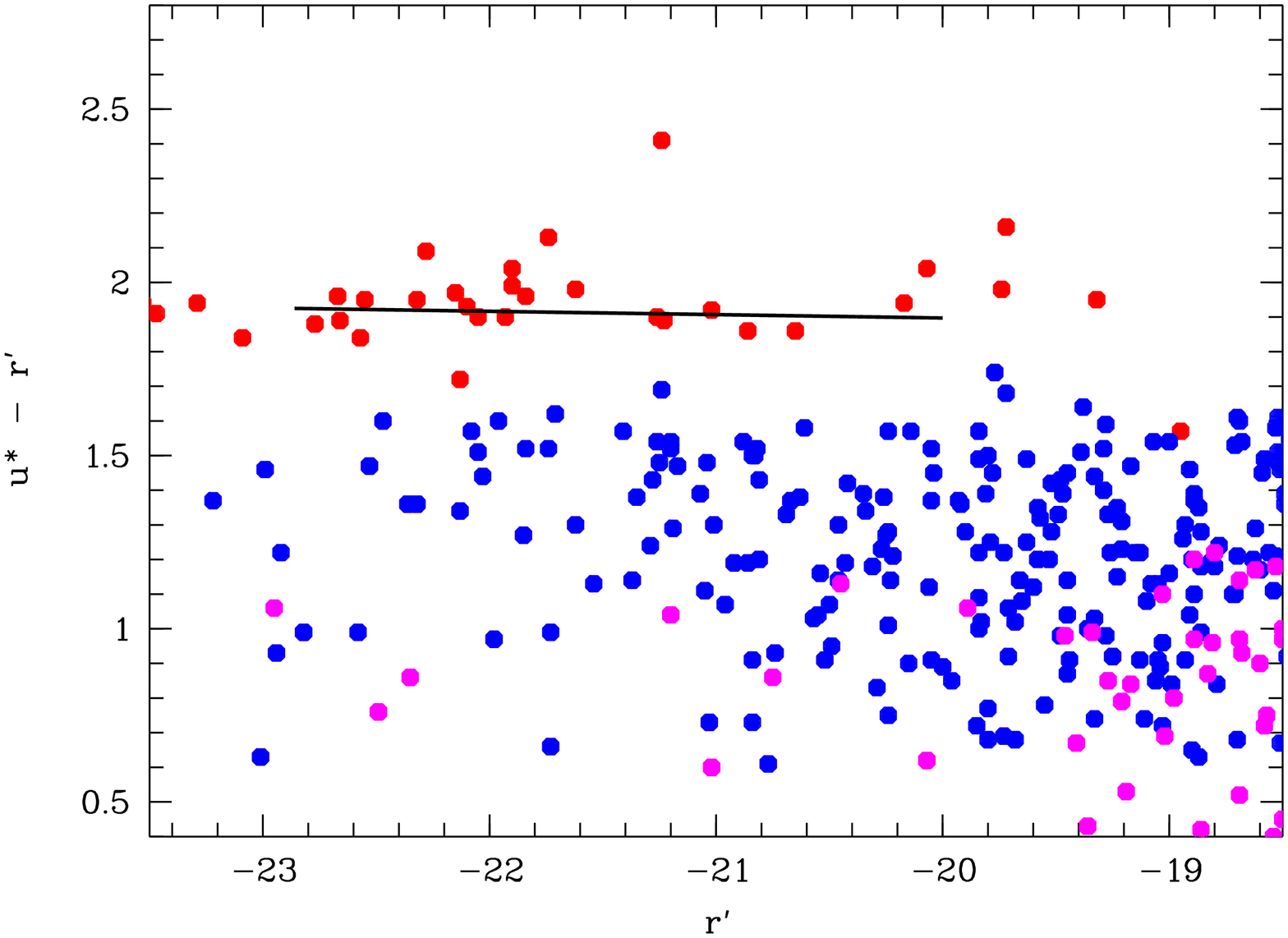,width=8cm,angle=270}}
\mbox{\psfig{figure=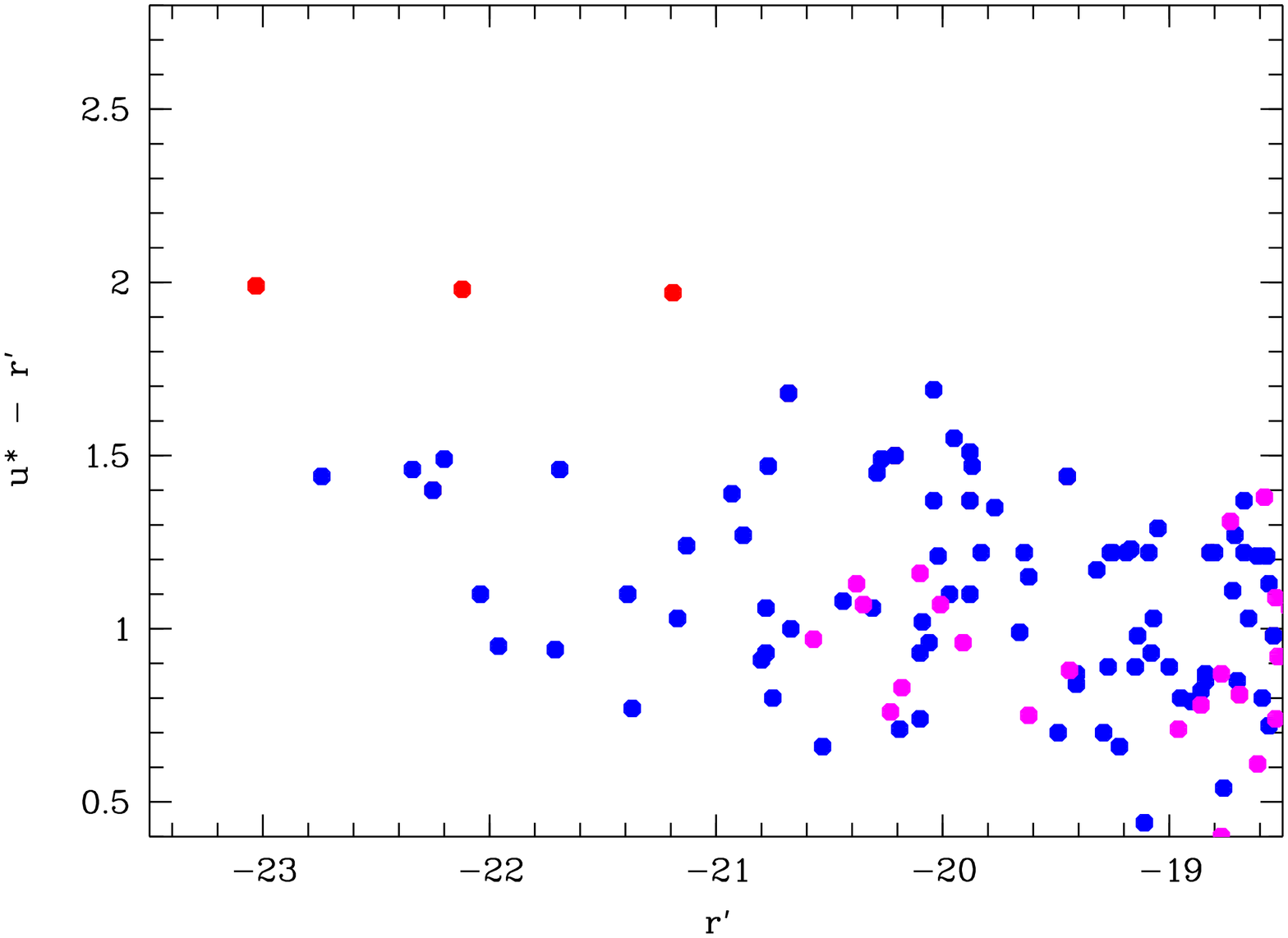,width=8cm,angle=270}}
\caption[]{u*-r' versus r' with red dots being T$\leq$21 galaxies (early types),
blue dots being 58$\geq$T$\geq$21 galaxies (late types), and pink dots being 
starburst galaxies. From top to bottom, figures are for the $most~massive$, the 
$massive$, the $moderately~massive$, and the z$\leq$1 C0 clusters. Absolute magnitude 
computations are based on photometric redshifts. Black continuous lines are the RSs 
computed with T$\leq$21 galaxies, except for C0 clusters where we had not enough available 
early type galaxies.}
\label{fig:fig10}
\end{figure}

In order to study larger samples and detect possible weak effects, we used 
photometric redshifts to define a cluster membership, and compute absolute magnitudes and colors as provided by 
the CFHTLS data. Given its photometric redshift, a galaxy was assigned to a cluster when closer than 500 kpc 
from the cluster center and at less than 0.08 from the cluster redshift. This corresponds to the values quoted 
in Table~\ref{tab:data} for cluster galaxies. We then were able to search for RSs in the $most~luminous$, the $luminous$, 
and the $moderately~luminous$ clusters.
Selecting all available clusters in these three categories Fig.~\ref{fig:fig10} 
clearly shows red sequences in each case. They all are consistent with a u*-r' color of 1.9,
the most massive clusters exhibiting the more negative RS slope (computed with $T\leq21$ galaxies). 
On the contrary, the C0 clusters (no X-ray detection) only exhibit
a very low number of early type galaxies (but still consistent with u*-r'$\sim$1.9). These
structures therefore appear as quite young structures, with modest early type galaxy populations.

\begin{figure}
\centering
\mbox{\psfig{figure=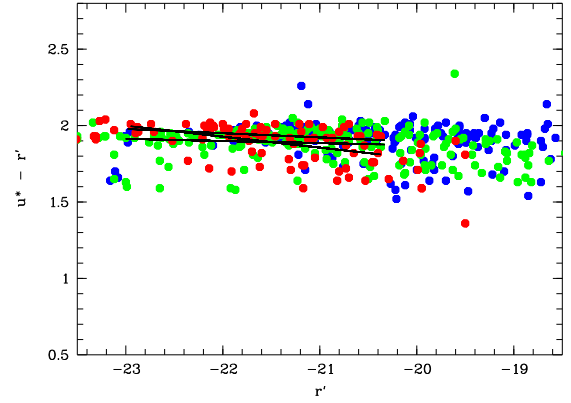,width=8cm,angle=270}}
\caption[]{u*-r' versus r' for T$\leq$21 galaxies (early types). Red dots: z=]0.5,0.8],
green dots: z=]0.3,0.5], and blue dots: z=[0.,0.3]. Black continuous lines are the RSs computed with 
 T$\leq$21 galaxies.}
\label{fig:fig10various}
\end{figure}

However, we are merging in Fig.~\ref{fig:fig10} clusters with quite different redshifts and
evolutionary effects could play an important role. We therefore selected only the $luminous$
clusters (the only category providing enough clusters) and we divided this population in 3 different
redshifts bins ($\leq$0.3, ]0.3,0.5], and ]0.5,0.8]) in Fig.~\ref{fig:fig10various}.
This figure only shows $T\leq21$ galaxies (early types). RSs appear very similar, with the most
negative slope occuring for z=]0.5,0.8] clusters. If evolutionary effects are present, they are
therefore rather weak, besides the most distant clusters appearing to have the most negative 
RS slope (-0.069$\pm$0.017). This is consistent with the slope computed for the $most~luminous$ 
clusters (which are also nearly all at redshift greater than 0.5):  -0.052$\pm$0.015.

\begin{figure}
\centering
\mbox{\psfig{figure=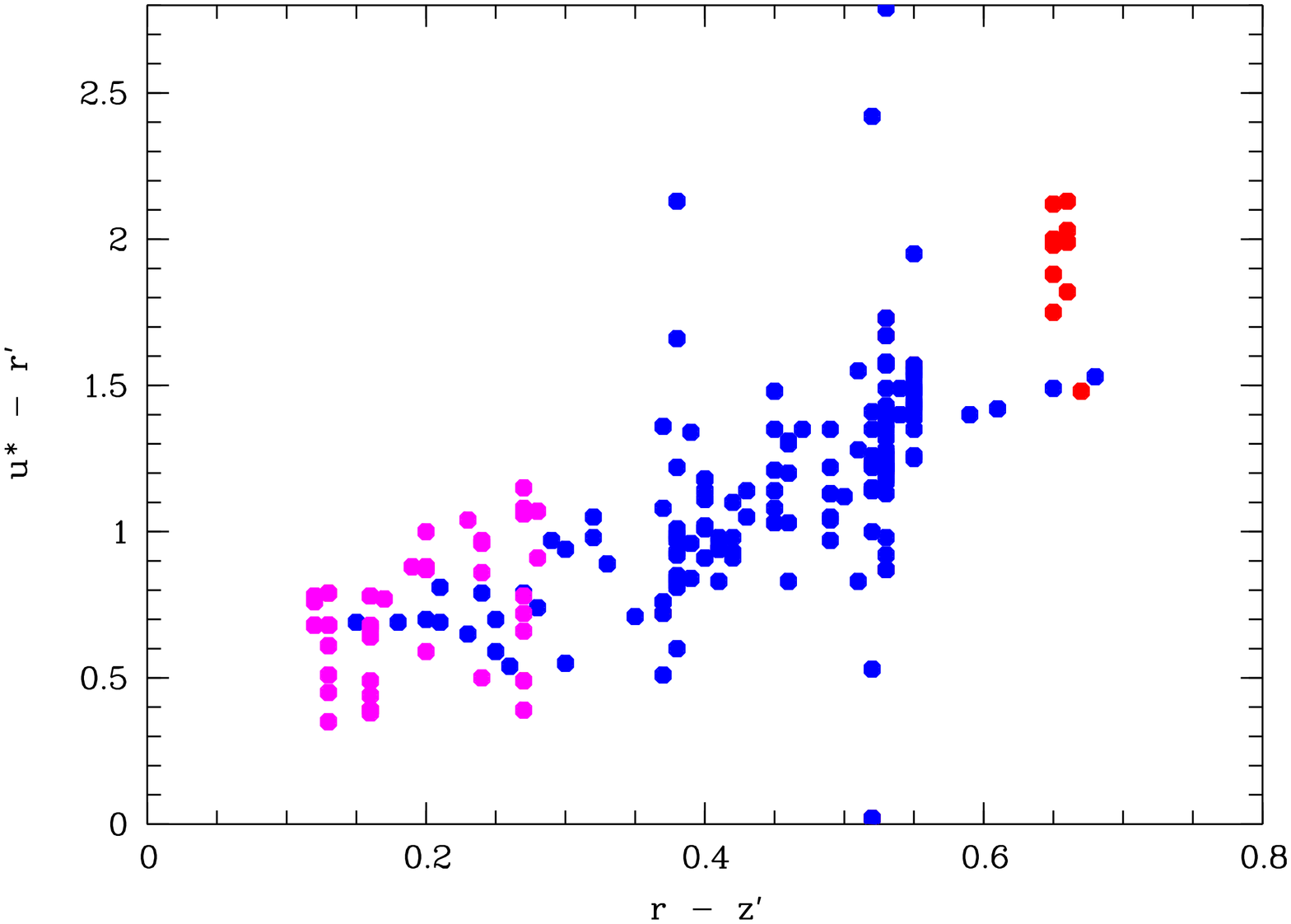,width=8cm,angle=270}}
\mbox{\psfig{figure=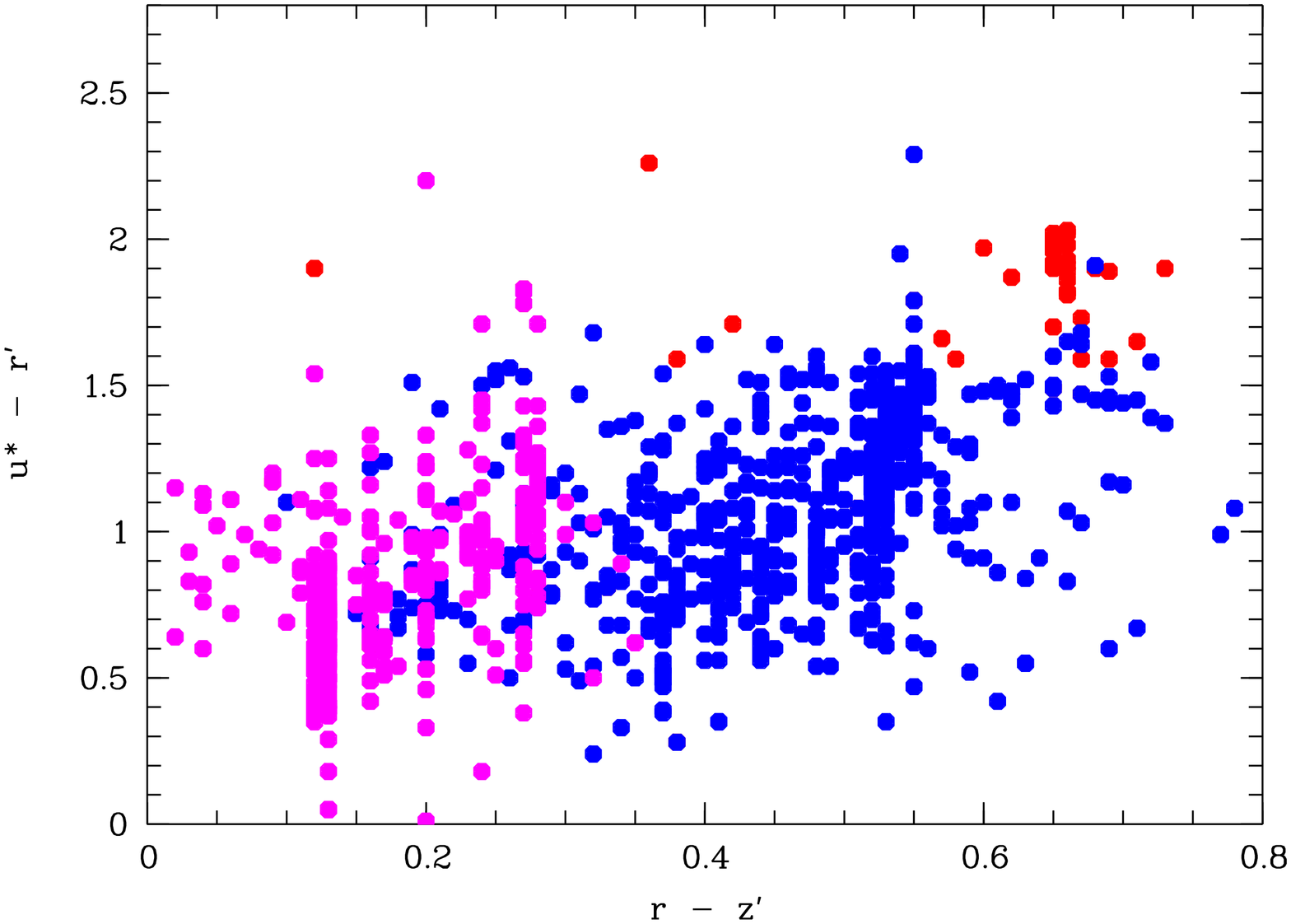,width=8cm,angle=270}}
\mbox{\psfig{figure=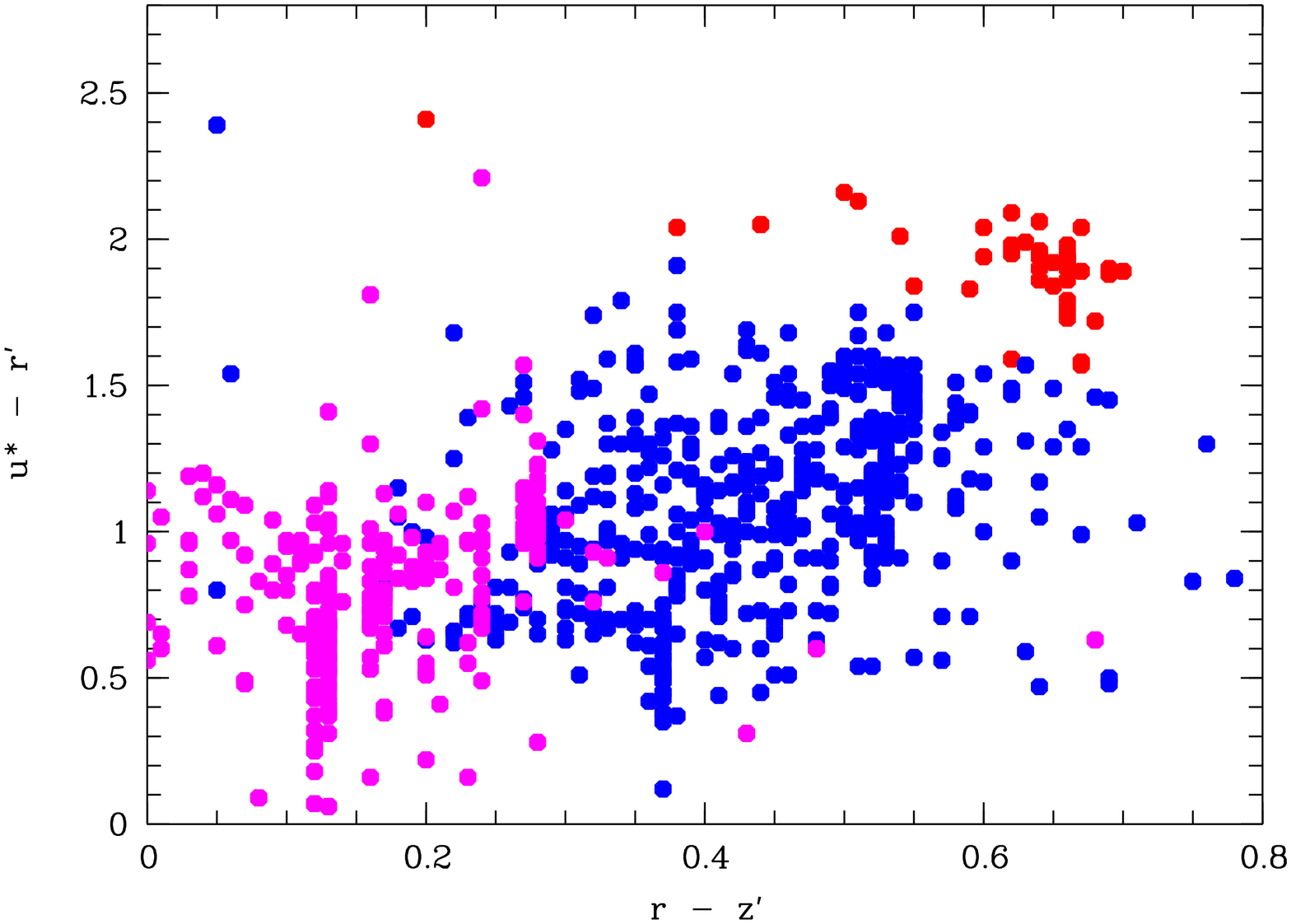,width=8cm,angle=270}}
\caption[]{u*-r' versus r'-z  diagrams  with red dots being T$\leq$21 galaxies (early types),
blue dots being 58$\geq$T$\geq$21 galaxies (late types), and pink dots being 
starburst galaxies. The upper figure is for the $most~luminous$, the middle figure is 
for the $luminous$, and the lower figure is for the $moderately~luminous$ clusters. }
\label{fig:fig11}
\end{figure}

It could be argued that the use of photometric redshifts could introduce a bias
due for example to SEDs not adapted to high density regions. In order to check the 
previous results, we therefore simply draw u*-r' versus r'-z' color-color diagrams 
for the same sets of clusters. 
Fig.~\ref{fig:fig11} shows that in both cases, early types still occupy well defined
loci in the color-color space, confirming the existence of an old galaxy
population in these cluster classes. 

We therefore confirm that both massive and less massive X-ray structures in our sample
exhibit quite similar red sequences, making them overall quite old structures. Non X-ray clusters
are probably minor structures with a poor spectral early type population. 
 
Fig.~\ref{fig:fig10} also shows a slightly larger percentage of starburst galaxies (as 
determined during the photometric redshift computation process: see Coupon et al., 2009) 
in low luminosity clusters. C0, $moderately~luminous$, and $luminous$ clusters exhibit 20$\%$ more
starburst galaxies compared to the $most~luminous$ clusters. This is an expected behaviour, for 
example in good qualitative agreement with Urquhart et al. (2010).

 \subsection{Luminosity Functions}

 In the same spirit, we checked whether our structures behave as genuine clusters or
 groups concerning their galaxy luminosity functions. For a detailed study 
 of the individual XMM-LSS C1 cluster luminosity functions, we refer the reader 
 to Alshino et al. (2010). We computed luminosity functions using
 galaxies within the cluster bins (according to photometric redshifts). The Schechter
 function fitting was performed allowing a constant background to take into
 account galaxies included in the photometric redshift slice but not part of
 the clusters. 

 Selecting all clusters (C1+C2+C3+C0), stacking their luminosity functions, and only limiting
 absolute magnitude to i'$\leq$-17.5 in order to not be too affected by incompleteness,
 we got a best fit of a Schechter function with alpha = -1.15$\pm$0.09 and M*i'=-23.8$\pm$0.8. 
 This is consistent within error bars with the estimates of Alshino et al. (2010) at 
 z$\sim$0.3. Playing the same game with the C3 clusters, we get a slightly shallower Schechter 
 fit: alpha = -0.96$\pm$0.14 and M*i'=-22.1$\pm$0.6. 

 If we use the luminosity categories, we can compute similarly Schechter 
 fits for the $luminous$ and the $moderately~luminous$ clusters ($most~luminous$
 clusters are too distant and therefore undersampled toward the faint magnitudes, and 
C0 and C2 clusters are not numerous enough).
 We get alpha = -1.1$\pm$0.03 and M*i'=-23.4$\pm$0.3 for the $moderately~luminous$
 and we get alpha = -1.1$\pm$0.03 and M*i'=-23.2$\pm$0.2 for the $luminous$ clusters.

The fitted Schechter
 functions are in agreement with those of bona fide clusters at similar depth
 (e.g. Lumsden et al. 1997). Slopes are also similar within error bars between 
 all cluster classes. C1 clusters seem to exhibit however, brighter $M_*$ than C3
 clusters, in good agreement with the fact that C1 clusters would be older and
 more massive systems than C3 clusters.

\section{Peculiar structures in the XMM-LSS}

\subsection {Distant cluster candidates}

Several structures with redshifts $\sim$1  or greater have already been
found in Class 1 (Pacaud et al 2007) or 2 (Bremer et al. 2006). 
Some other candidates appear among the C2's (e.g. XLSS J022756.3-043119 at z $\sim$1)  and 
the C0's (022550.4-044500 at z $\sim$1.53). 
This last structure (Fig.~\ref{fig:superdistant}) is just below the
X-ray detection limit. It has an extension of $\sim$13 arcsec and its extension maximum 
likelihood is $\sim$10. We note that the measured flux is 0.2$\pm$0.1 10$^{-14}$ erg/s/cm$^2$. At z=1.53 
and for a temperature of 1.5keV, this would lead to an X-ray luminosity of 8.6e+43 ergs/s.

The weakness of the evidences for an X-ray detection leads us, however, to classify this source 
as C0 and then to investigate it from the optical side.
The regular CFHTLS
photometric redshifts (based on u*g'r'i'z' magnitudes) are not well suited to study 
this potential structure
because of a lack of near infrared photometric bands. This candidate is however 
included in the WIRDS survey (near infrared imaging from CFHT-WIRCAM). Photometric 
redshifts have been computed combining these near infrared data (McCracken et al., private
communication) and the CFHTLS deep magnitudes. Fig.~\ref{fig:superdistant} shows
a clear concentration of z=[1.43;1.63] galaxies inside the XMMLSS contours. We therefore
may have detected one of the most distant known clusters of galaxies. A near infrared 
spectroscopic follow up of this candidate is however mandatory in order to confirm
the nature of this very weak X-ray source.

\begin{figure}
\centering
\caption[]{A distant cluster candidate at z = 1.53. Large red circle
is a 500 kpc radius circle. Blue circles are galaxies with spectroscopic redshifts
outside the z=[1.52;1.54] interval. The two magenta squares are the two known 
spectroscopic redshifts inside the z=[1.52;1.54] interval. Small red circles are
the near-infrared-based photometric redshifts inside the z=[1.43;1.63] interval. White
contours are the XMM-LSS contours.}
\label{fig:superdistant}
\end{figure}

\subsection{Structures with discrepant optically and X-ray contents}

XLSSC 000 is a C0 structure not detected in the X-rays. Its velocity dispersion is
however relatively large (435$\pm$88 km/s). The Serna-Gerbal analysis 
does not detect any sign of substructures with the 11 known spectroscopic redshifts,
so this velocity dispersion does not appear as obviously biaised. The photometric 
redshift distribution  also presents excesses at the structure redshift.  This 
structure is finally populated with a significant number of early types galaxies: among the 21
 objects within the z=0.49 photometric redshift slice, 9 have type T$\leq$10. 
The optical content is therefore similar to what we could expect if considering
a massive cluster.
This case with clear discrepancies between X-ray and optical content remains 
quite puzzling and both deeper X-ray observations and additional spectroscopic followup are 
required to explain the observed behaviour.

We also have detected a prominent X-ray structure which is much less evident in optical
and which could be a fossil group (XLSSU J021754.6-052655). Described for example in
Jones et al. (2003) or Mendes de Oliveira et al. (2006 and references
therein), these structures are considered as the ultimate stage of group
evolution: the nearly complete fusion of all the bright and
intermediate magnitude galaxies of the group into a single bright
galaxy. The resulting galaxy is brighter than the second remaining
group galaxy (within half the projected virial radius) by at least 2
magnitudes (in the R band). However, the extended X-ray gas envelope 
is still present and more luminous than 10$^{42}~$h$_{50}^{-2}$ erg~s$^{-1}$ 
(Jones et al. 2003). 
The origin of these structures is however being still widely debated. They could
find their origin in the small impact parameter of $L \sim L_*$ galaxies
travelling along filaments (e.g. D'Onghia et al. 2005), or simply in their 
highly isolated status (e.g. Adami et al. 2007a) so that
   no galaxies will then have fallen into them lately. 

In our survey, XLSSU J021754.6-052655 (classified as C2) is quite similar to
such fossil groups. Fig.~\ref{fig:fg2022} shows the field covered by this galaxy
structure. The X-ray source is clearly extended. Available spectroscopic
redshifts only show 2 galaxies at the structure redshift which are only slightly
too bright to satisfy the 2 magnitudes criteria (one is satisfying the
criterion in i' and z' band). Photometric redshifts from Coupon et al. (2009)
also exhibit only two other similar galaxies at less than 0.15 
from the structure redshift. Considering error bars on magnitude, the
magnitude difference between the brightest galaxy and the second brightest
object could be consistent with the requested 2 magnitudes gap at the
3-$\sigma$ level. We therefore conclude that this object is similar to 
the structure described in Ulmer et al. (2005) and is very close to the fossil 
group status.

\begin{figure}
\centering
\caption[]{CFHTLS i' band image of the XLSSU J021754.6-052655 XMM-LSS source. White contours are
XMM X-ray emission. Pink squares are galaxies with a spectroscopic redshift
inside the structure. Red circles are galaxies with a photometric redshift
at less than 0.15 from the structure redshift. Large symbols (circles or
squares) are galaxies not satisfying the 2 magnitudes criteria in g', r', i',
or z' bands. For these objects we also give the magnitude difference with the
brightest galaxy in g'/r'/i'/z'.}
\label{fig:fg2022}
\end{figure}

\begin{figure}[hbt] 
\centering \mbox{\psfig{figure=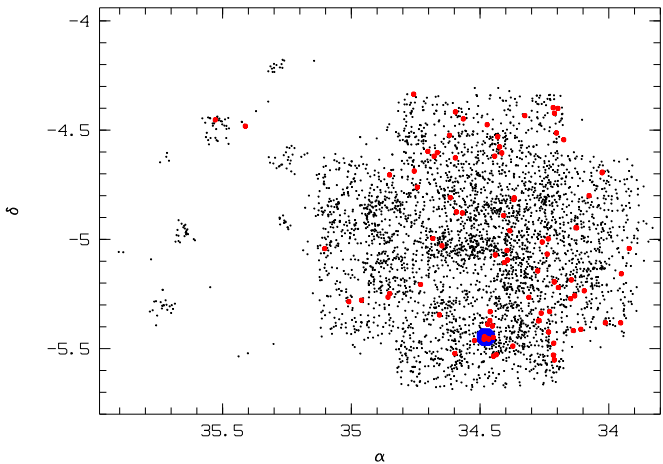,width=9.25cm,angle=270}}
\caption[]{$\alpha$,$\delta$ map of the immediate vicinity of XLSS J021754.6-052655. 
Black dots are all galaxies with a known spectroscopic redshift in a
1.5deg radius region (large blue filled circle). Red
dots are galaxies in the redshift range [0.241,0.261].} 
\label{fig:FG}
\end{figure}

We investigate if this group is the dominant structure of its
cosmological bubble (similarly to Adami et al. 2007a).
For this, we selected all known spectroscopic redshifts in
the range [0.241,0.261] and at less than 1.5deg from the group (about
20 Mpc at the structure redshift, close to the average size of known voids:
e.g. Hoyle \& Vogeley 2004). Contrary to the results of Adami et al. (2007a),
our group does not appear as an isolated structure (Fig.~\ref{fig:FG}). The ratio of
galaxies with a spectroscopic redshift inside and outside the range [0.241,0.261] is not
significantly different when considering the 1.5deg region or the complete
spectroscopic sample. 

\section{Conclusions}

Starting from known XMM-LSS sources, we considered
75 of them for which at least two spectroscopic redshifts  
were available within the X-ray isophotes. We then generated a catalog of 59
groups or clusters of galaxies in the z=[0.05;1.53] redshift range associated with an
X-ray source as well as 7 other real structures for which X-ray association is not clear.
Finally, 11 redshift structures (named C999)  detected along the various lines of sight were 
detected in addition of the main systems and are listed in Appendix. 
In 3 cases the X-ray sources are in fact associated with QSO's
identified from their optical spectra. 

The assessment of the clusters and groups  as actual massive structures
has been based on various spectroscopic data (including PI observations)
associated to photometric data from the CFHTLS T0004 release (when available) 
and some PI data. The analysis (without a-priori knowledge of
their X-ray class) of the optical lines of sight centered on the X-ray emission was
based on criteria such  as compactness in redshift space (spectroscopic and
photometric),  and  significant excess in galaxy density obtained  within
photometric redshift slices and  final visual inspection.  

All the detected systems exhibit "bona fide" clusters or groups optical
properties in terms of red sequence, color-color clumping, luminosity
function, and morphological segregation.  Considering X-ray luminosity classes
does not change the results.
From the X-ray and optical  properties of the structures now  associated with
the XMM extended sources, the C1 clusters can be considered in most cases as
relatively nearby, X-ray bright and  optically rich and regular (no
sub-clustering) clusters, while C3's appear faint and poor at the same
redshift and quite rich at high redshift. C2's are a mix with the exception 
of some distant possible candidates.
Finally, looking at larger scales using  the CFHTLS-W, these clusters statistically
appear as clear nodes of the galactic cosmic web, reinforcing therefore their 
true existence. 
The full sample of X-ray clusters with associated optical spectroscopic data 
is available via the L3SDB database (http://l3sdb.in2p3.fr:8080/l3sdb/).
The optical images as well as the details of the redshift determination for all 
clusters presented in this article will also be publicly available at this place.

Finally, we investigated the photometric redshift precision in our sample as a function
of the environment and of the galaxy spectral types (see appendix). We show for example that 
the galaxy photometric redshift accuracy is degraded in the most massive clusters for early and late
type galaxies.

\begin{acknowledgements}
The authors thank R. Bielby, J. Coupon, Y. Mellier, and H.J. McCracken for help.
AD, TS and JS acknowledge support from the ESA PRODEX Programme
"XMM-LSS", from the Belgian Federal Science Policy Office and from
the Communaut\'e francaise de Belgique - Actions de recherche concert\'ees -
Acad\'emie universitaire Wallonie-Europe. HQ acknowledges the support of 
FONDAP Center for Astrophysics 15010003. GG is supported by FONDECYT 1085267.
The authors thank the referee for useful and constructive comments.
\end{acknowledgements}

\appendix

\section{Photometric redshifts in dense environments}
  
A by-product of the present paper is the test of photometric redshift precision in
dense environments. Photometric redshift technique is widely used for
several cosmological purposes, and is mainly based on synthetic energy
distributions (SEDs hereafter) fits to observed magnitudes. The
available SEDs in the literature are however mainly selected in low density
environments, outside clusters. Applying these SEDs to cluster galaxies is
then potentially problematic. Several papers (e.g Adami et al. 2008) seem to show various  
photometric uncertainties as a function of the 
galaxy spectral type in these dense environements. If confirmed, this could be due 
in massive structures 
to environmental effects driving peculiar color galaxy
evolutions. Degeneracies could then be induced between photometric redshift value and galaxy
spectral type when applying classical photometric redshift codes as LePhare
or HyperZ (Bolzonella et al. 2000). However, these
tendencies are still based on very sparse samples for clusters of galaxies and
before embarking in the very demanding task of building cluster-dedicated
SEDs, we have to put on a firmer ground the photometric redshift uncertainty
variation as a function of the environment and of the galaxy spectral
type. 

The XMM-LSS survey offers such a unique opportunity, both providing X-ray and optical
characterizations of the clusters, and photometric redshift informations from
the CFHTLS. We selected all spectroscopic redshifts included in the present
clusters and located in the 500 kpc (radius) central area. This insures us to have
galaxies really located in the densest areas of the clusters. Then, we
extracted informations (photometric redshift itself and spectral type) from
the CFHTLS T0004 photometric redshift release. Finally, we considered separately 
clusters brighter than 10$^{44}$ erg/s, between 10$^{43}$ and 
10$^{44}$ erg/s, and fainter than 10$^{43}$ erg/s. We also considered the C0 class, 
acting as the low mass cluster category (we remind
that these structures are real but without clear X-ray emission).

\subsection{General agreement between spectroscopic and photometric
redshifts}

\begin{figure}[hbt] 
\centering \mbox{\psfig{figure=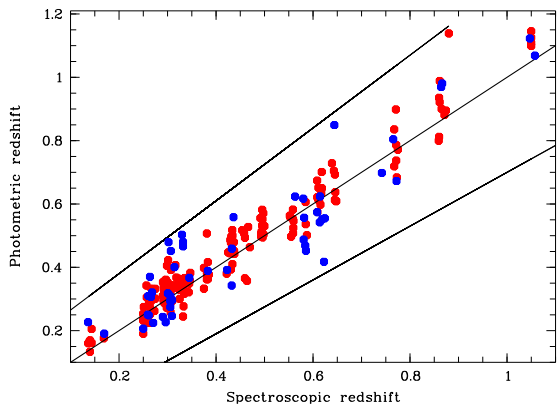,width=8.cm,angle=270}}
\caption[]{Photometric versus spectroscopic redshifts for the cluster galaxies in our
  sample. Black lines give the
  perfect relation of slope 1 and the $\pm$0.15$\times$(1+z) classical uncertainty (see
  e.g. Ilbert et al. 2006). Red filled circles are early spectral type
  galaxies and blue filled circles are late spectral type
  galaxies (see text).} 
\label{fig:global}
\end{figure}

We first checked that the general agreement between spectroscopic and photometric
redshifts was acceptable inside clusters of galaxies. Fig.~\ref{fig:global} shows a good
agreement. Selecting a priori galaxies with a $|zphot - zspec|\leq(0.15\times(1+zspec))$, 
the whole cluster galaxy sample exhibits a $\sigma$ of 0.06. This is only slightly
larger than the estimates of Coupon et al. (2009) for the whole CFHTLS W1 field.
This shows that from a general point of view, CFHTLS T0004 photometric redshifts are not
clearly worst in clusters than in the field. We have now to investigate
in more details the behaviour of the photometric redshift uncertainty as a
function of the galaxy spectral type and as a function of the cluster characteristics. 

\subsection{Photometric redshifts in dense environments and galaxy spectral types}

We redid the previous analysis splitting our samples into early and late type
galaxies and considering the C0, the $most~luminous$, the $luminous$, and the 
$moderately~luminous$ clusters. Table~\ref{tab:data} gives the values of $\sigma$ (computed in the 
same way as in the previous subsection) in these different cases.

\begin{table}
\caption{$\sigma$ between photometric and spectroscopic redshifts as a
  function of the environment and of the galaxy spectral type. The last column 
indicates the percentage of spectral types $ T \leq$ 21 galaxies.}
\begin{center}
\begin{tabular}{ccccc}
\hline
Cluster class & Early & Late & $\%$ of early types\\
\hline
$most~luminous$ & 0.081   & 0.092 & 69$\%$ \\
$luminous$ & 0.036   & 0.082 & 81$\%$ \\
$moderately~luminous$ & 0.047   & 0.069 & 72$\%$ \\
C0 & 0.043   & 0.064 & 75$\%$ \\
Global cluster sample & 0.048   & 0.096 & 76$\%$ \\  
\hline
\end{tabular}
\end{center}
\label{tab:data}
\end{table}

We first detect a clear tendency to have higher uncertainties in photometric
redshift calculations in the $most~luminous$ clusters. Second, late type galaxies 
in $luminous$ and $moderately~luminous$ clusters (as well as in C0 clusters) also
exhibit higher uncertainties than early type galaxies, by a factor of 2 in $luminous$ 
clusters and by 50$\%$ in $moderately~luminous$ and C0 clusters. This behaviour was 
already detected in Guennou et al. (2010). 

This can be explained if galaxies were undergoing peculiar evolutions in clusters 
of galaxies, depending of the mass of the considered clusters, making them different 
from field galaxies. This would occur for all galaxy types in the most massive clusters, while
less massive clusters would only affect late type galaxies. These
various environments do not seem to strongly affect the percentage of early type galaxies 
which stays high anyway (see Table~\ref{tab:data}). A finer analysis shows however, as expected, 
a regular increase of the mean type of $ T \leq$ 21 galaxies, from 1.4 for the $most~luminous$
clusters, to 2.2 for the $luminous$ clusters, and finally to 3.0 for the $moderately~luminous$
clusters. 

As a conclusion, we can then say that photometric redshift values are
globally correct in clusters of galaxies of the present sample (as compared to field
environments). However, all galaxies in the most massive clusters and late type galaxies
in all other clusters have their photometric redshift uncertainty increased by a factor of 
50 to 100$\%$. Depending on the science goals, this can significantly affect the cluster 
population definition by photometric redshift criteria, for example for galaxy luminosity
function purposes. It would therefore be useful to create cluster-dedicated spectroscopic SEDs.

\section{Additional redshift structures}
  
 As a bonus of the general cluster detection process, for a given line of sight, other real 
 galaxy groups are detected
 besides the ones associated with the X-ray emission (C999: see Table~\ref{tab:cand2}). 
 So, if the identification with an optical group would appear wrong in the future, or if
 more data become available, trace is kept to re-examine other possibilities.

\begin{table}
\caption{Same as Table~\ref{tab:cand1C2} but for other real groups (C999) detected
along the lines of sight.}
\begin{center}
\begin{tabular}{cccccccccc}
\hline
XLSSC & PH & RA & DEC & N & ZBWT & ERRZ & SIG &
ERR \\
    &        &      & deg & deg &   &      & km/s &
km/s \\
\hline
065     & 1 & 34.245 & -4.821 & 3 & 0.138 &     &  &   \\
039     & 0 & 35.098 & -2.841 & 3 & 0.183 &     &  &    \\
044     & 1 & 36.141 & -4.234 & 11 & 0.317 & 0.001 & 410 & 87\\
-   & 1 & 36.424 & -4.410 & 4 & 0.142 &     &  &    \\
-   & 1 & 36.424 & -4.410 & 4 & 0.632 &     &  &    \\
-   & 1 & 36.424 & -4.410 & 3 & 0.915 &     &  &    \\
-   & 1 & 36.460 & -4.750 & 3 &     &     &  &    \\
-   & 1 & 36.698 & -4.241 & 3 & 0.210 &     &  &    \\
-   & 1 & 36.698 & -4.241 & 3 & 0.432 &     &  &    \\
-   & 1 & 36.698 & -4.241 & 3 & 0.705 &     &  &    \\
013     & 1 & 36.858 & -4.538 & 9 & 0.254 & 0.001 & 346 & 75\\
\hline
\end{tabular}
\end{center}
\label{tab:cand2}
\end{table}

\end{document}